\documentclass[12pt]{article}
\usepackage{amsmath,bbold,hyperref}
\usepackage[T1]{fontenc}	% allows copy-paste of special characters from the pdf
\usepackage[utf8]{inputenc}
\usepackage{xcolor}
\usepackage{braket}

\newcommand{\be}{\begin{equation}}
\newcommand{\ee}{\end{equation}}
\newcommand{\bea}{\begin{eqnarray}}
\newcommand{\eea}{\end{eqnarray}}
\newcommand{\beas}{\begin{eqnarray*}}
\newcommand{\eeas}{\end{eqnarray*}}

\newcommand{\G}[1]{\Gamma\left(#1\right)}

\newcommand{\hyp}[1]{{}_2F_1\left(#1\right)}

\begin{document}
\begin{titlepage}

\begin{center}

{\Large Bulk reconstruction for spinor fields in AdS/CFT}

\vspace{12mm}

\renewcommand\thefootnote{\mbox{$\fnsymbol{footnote}$}}
Valentino F.\ Foit${}^{1}$\footnote{foit@nyu.edu},
Daniel Kabat${}^{2}$\footnote{daniel.kabat@lehman.cuny.edu},
Gilad Lifschytz${}^{3}$\footnote{giladl@research.haifa.ac.il}

\vspace{4mm}

${}^1${\small \sl Center for Cosmology and Particle Physics} \\
{\small \sl New York University, New York, NY 10003, USA}

\vspace{2mm}

${}^2${\small \sl Department of Physics and Astronomy} \\
{\small \sl Lehman College, City University of New York, Bronx, NY 10468, USA}

\vspace{2mm}

${}^3${\small \sl Department of Mathematics and} \\
{\small \sl Haifa Research Center for Theoretical Physics and Astrophysics} \\
{\small \sl University of Haifa, Haifa 31905, Israel}

\end{center}

\vspace{8mm}

\noindent
We develop the representation of free spinor fields in the bulk of Lorentzian anti-de Sitter space in terms of smeared operators in the dual conformal field theory.
To do this we expand the bulk field in a complete set of normalizable modes, work out the extrapolate dictionary for spinor fields, and show that the bulk field can be reconstructed
from its near-boundary behavior.  In some cases chirality and reality conditions can be imposed in the bulk.  We study the action of the CFT modular
Hamiltonian on bulk fermions to show that they transform with the expected spinor Lie derivative, and we calculate bulk--boundary two-point functions starting from CFT correlators.

\end{titlepage}
\setcounter{footnote}{0}
\renewcommand\thefootnote{\mbox{\arabic{footnote}}}
%%%%%%%%%%%%%%%%%%%%%%%%%%
\section{Introduction}
%%%%%%%%%%%%%%%%%%%%%%%%%%
The AdS/CFT correspondence \cite{Maldacena:1997re} relates a theory of gravity in anti-de Sitter space to a large-$N$ conformal field theory on the boundary.  
One aspect of the
correspondence is that, perturbatively in the $1/N$ expansion, bulk quantum fields can be expressed as CFT operators.  In the large $N$ limit the bulk fields are free and can
be written as smeared single-trace operators in the CFT.  
In the $1/N$ expansion interactions can be taken into
account by adding multi-trace corrections.
This has been established explicitly for bulk scalars as well as for bulk fields with integer spin \cite{Hamilton:2006fh,Kabat:2011rz,Kabat:2015swa}.\footnote{For reviews of bulk reconstruction see \cite{DeJonckheere:2017qkk,Harlow:2018fse}.}

To fill in a somewhat neglected corner of the AdS/CFT correspondence, in this paper we develop the CFT representation of bulk fields with spin $1/2$.  Bulk fermions have been studied before \cite{Henningson:1998cd,Mueck:1998iz}, generally in a Euclidean
setting where the CFT is deformed by adding sources \cite{Gubser:1998bc,Witten:1998qj}.  By contrast we work in Lorentzian signature and we do not introduce sources in the CFT.  Instead we construct on-shell bulk fields describing
normalizable fluctuations about the AdS vacuum.  According to the extrapolate dictionary \cite{Klebanov:1999tb,Harlow:2011ke}, as a bulk point approaches the boundary such fields go over to local operators in the CFT.  Thus at leading
large $N$, where the bulk fields are free, our goal is to represent an on-shell bulk field as a smeared operator in the CFT.  To do this we will follow the approach developed in \cite{Hamilton:2006fh,Kabat:2012hp} for fields with integer spin and extend it to fields with spin $1/2$.

The CFT operators that we construct should have all the properties of free bulk fields in AdS.  As one test of this, we will show that they can be used to recover bulk -- boundary correlators from the CFT.  As another test we study their behavior under CFT modular flow.  Modular flow in the CFT is dual to modular flow in the bulk \cite{Jafferis:2015del}, which in the vacuum reduces to flow along an AdS isometry.  Bulk fields should transform with the appropriate Lie derivative, and we check that this is indeed the case.

An outline of this paper is as follows. In Section \ref{sect:modesum} we develop the representation of free bulk fermions in the CFT.  We do this by expanding the bulk field in normalizable modes. There are two types of modes we consider, normalizable in different mass ranges.  Moreover spinor representations make a sharp distinction between even- and odd-dimensional AdS, so we analyze all these
cases separately. 
In Section \ref{sect:correlators} we recover bulk -- boundary correlators from the CFT by applying the smearing functions to a CFT two-point function.
In Section \ref{sect:modular} we study the action of the CFT modular Hamiltonian on the bulk fermions we have constructed and in Section \ref{sect:conditions} we consider the various
possible chirality and reality conditions that can be imposed on a bulk fermion.
The appendices contain some supporting material. Appendix \ref{appendix:spin} gives our conventions for spin matrices.  In Appendix \ref{appendix:BF} we study the conditions for a bulk fermion to be normalizable and we discus permissible boundary conditions for
fermions in AdS.  We review the analog of the Breitenlohner-Freedman (BF) bound \cite{Breitenlohner:1982jf} for fermions and note that there is a range of masses $- {1 \over 2} < m < {1 \over 2}$ where inequivalent quantizations are possible.  Appendix \ref{appendix:spinor} reviews spinor representations in general dimensions.

%%%%%%%%%%%%%%%%%%%%%%%%
\section{Smearing functions from mode sums\label{sect:modesum}}
%%%%%%%%%%%%%%%%%%%%%%%%
In this section we derive the smearing functions that let us represent a Dirac fermion $\Psi(T,{\bf X},Z)$ in AdS${}_{d+1}$ in terms of a spinor field $\psi(t,{\bf x})$
in CFT${}_d$.  We do this using a bulk mode expansion, following the approach developed for scalar fields in \cite{Hamilton:2006fh}.  Compared to scalar fields, the main novelty
is that the fermion representations are rather different depending on whether $d = ($number of spacetime dimensions in the CFT$)$ is even or odd.
\begin{itemize}
\item
If $d$ is even then the bulk and boundary Dirac matrices have  the same size.  We will see that a Dirac fermion in the bulk is dual (in the sense of the extrapolate dictionary) to a chiral fermion on the boundary.  The smearing functions will let us reconstruct a bulk Dirac fermion from a boundary chiral fermion.
\item
If $d$ is odd then a bulk Dirac fermion has twice as many components as a boundary Dirac fermion.  In this case we will see that a Dirac fermion in the bulk is dual (in the sense of the extrapolate dictionary) to a Dirac fermion on the boundary.  The
smearing functions will let us reconstruct a bulk Dirac fermion from a boundary Dirac fermion.
\end{itemize}
Note that in both cases half of the spinor degrees of freedom disappear as one goes to the boundary.

By way of outline, below we'll first set up the Dirac equation in general AdS.  Then we'll study the extrapolate dictionary and construct smearing functions, first in odd-dimensional AdS then in even-dimensional AdS.

%%%%%%%%%%%%%%%%%%%%%%%%
\subsection{Preliminaries\label{sect:preliminaries}}
%%%%%%%%%%%%%%%%%%%%%%%%
We start with the metric for the Poincar\'e patch of Lorentzian AdS${}_{d+1}$,
\begin{equation}
\label{Poincare}
ds^2 = {R^2 \over Z^2} \left(-dT^2 + \vert d\textbf{X} \vert^2 + dZ^2\right).
\end{equation}
Here $T, \textbf{X}, Z$ are bulk or AdS coordinates. Below, when we want to distinguish bulk and boundary coordinates, we will write the boundary coordinates as $t, {\bf x}$. From now on we set the AdS radius $R = 1$.

Our notation is as follows. $A, B$ are local Lorentz indices in the bulk
and $M, N$ are coordinate indices in the bulk; these capital indices range over all dimensions.  We will use lowercase letters $a, b$ to denote local Lorentz indices in the CFT directions (i.e.\ excluding the $Z$ direction)
and $m, n$ to denote bulk coordinate indices that exclude $Z$.  In Section \ref{sect:modular} we will use $i, j$ to denote local Lorentz indices that exclude both $T$ and $Z$.  The bulk fermion will be
denoted $\Psi$, with bulk Dirac matrices $\Gamma^A$, while the CFT fermion will be denoted $\psi$ with CFT Dirac matrices $\gamma^a$.

For a vielbein $e^A = e^A_M dx^M$ we take
\begin{equation}
\label{e}
e^A_M = {1 \over Z} \delta^A_M.
\end{equation}
These satisfy $ds^2 = e^A e^B \eta_{AB}$ where
$\eta_{AB} = \operatorname{diag}(-+\cdots+)$ is the Minkowski metric.  The corresponding spin connection
\begin{equation}
\omega^{AB} = \omega_M^{AB} dx^M
\end{equation}
is given by
\begin{eqnarray}
\nonumber
&& \omega^{ab} = 0 \\
\label{omega}
&& \omega^{aZ} = - \omega^{Za} = - e^a \\
\nonumber
&& \omega^{ZZ} = 0.
\end{eqnarray}
The spin connection has been chosen so that the torsion vanishes
\begin{equation}
\label{torsion}
de^A + \omega^A{}_B \wedge e^B = 0.
\end{equation}

Next we introduce a set of Dirac matrices $\Gamma^A$ satisfying $\lbrace \Gamma^A,\Gamma^B \rbrace = -2\eta^{AB} \mathbb{1}$.  We will be more explicit about the
form of these matrices below as it changes depending on whether $d$ is even or odd.  The action for a free Dirac fermion in AdS${}_{d+1}$ is
\begin{equation}
\label{action}
S = \int d^{d+1}x \sqrt{-g} \, \bar{\Psi} \left(i \Gamma^A e^M_A D_M - m \right) \Psi,
\end{equation}
where $\bar\Psi = \Psi^\dagger \Gamma^0$ is the Dirac adjoint, and the covariant derivative is
\begin{equation}
\label{SpinorD}
D_M = \partial_M - {1 \over 8} \omega_M^{AB} [\Gamma_A,\Gamma_B].
\end{equation}
The conventions that enter in this covariant derivative are discussed in Appendix \ref{appendix:spin}.  Given the spin connection (\ref{omega})
the covariant derivative has components
\begin{align}
\begin{split}
D_m &= \partial_m + {1 \over 2} e^a_m \Gamma_a \Gamma_Z \\
D_Z &= \partial_Z.
\end{split}
\end{align}
Using the fact that $\Gamma^a \Gamma_a = -d$, the Dirac operator in AdS is simply\footnote{We're slightly mixing notation: on the right-hand side we've made use of the
vielbein (\ref{e}), so the Dirac matrices $\Gamma^a$, $\Gamma^Z$ are in a local Lorentz basis while the derivatives $\partial_a$, $\partial_Z$ are in a coordinate basis.}
\begin{equation}
\label{DiracOp}
\Gamma^A e^M_A D_M = Z \Gamma^a \partial_a + Z \Gamma^Z \partial_Z - {d \over 2} \Gamma^Z.
\end{equation}

%%%%%%%%%%%%%%%%%%%%%%%%
\subsection{AdS${}_{\rm odd}$ / CFT${}_{\rm even}$\label{sect:AdSodd}}
%%%%%%%%%%%%%%%%%%%%%%%%
When the CFT spacetime dimension $d$ is even the bulk and boundary Dirac matrices have the same size, namely $2^{d/2} \times 2^{d/2}$.  To construct these matrices we begin with the CFT.
There we have a set of $d$ Dirac matrices $\gamma^a$ satisfying $\lbrace \gamma^a, \gamma^b \rbrace = - 2 \eta^{ab} \mathbb{1}$.  Since the spacetime dimension is even we also have a chirality
operator $\gamma$ which we take to be diagonal
\begin{equation}\label{gammachiral}
\gamma = \left(\begin{array}{cc} - \mathbb{1} & 0 \\ 0 & \mathbb{1} \end{array}\right).
\end{equation}
That is, we're using a chiral basis for the CFT Dirac matrices.  We'll write a boundary Dirac spinor in terms of its chiral components as $\psi = \left({\psi_- \atop \psi_+}\right)$.

The chirality operator satisfies $\lbrace \gamma, \gamma^a \rbrace = 0$ as well as $\gamma^2 = \mathbb{1}$.  This means the CFT Dirac matrices have the form
\begin{equation}\label{CFTevenDirac}
\gamma^a = \left(\begin{array}{cc} 0 & \sigma^a \\ \bar{\sigma}^a & 0 \end{array}\right),
\end{equation}
where the matrices $\sigma^a$, $\bar{\sigma}^a$ satisfy\footnote{As in four dimensions the overbar is just part of the notation for these matrices. It doesn't indicate complex conjugation.}
\begin{align}
\begin{split}\label{sigmaAlgebra}
\sigma^a \bar{\sigma}^b + \sigma^b \bar{\sigma}^a &= - 2 \eta^{ab} \mathbb{1} \\
\bar{\sigma}^a \sigma^b + \bar{\sigma}^b \sigma^a &= -2 \eta^{ab} \mathbb{1}.
\end{split}
\end{align}
For example in a 2-D CFT we could take $\sigma^a = (-i,i)$ and $\bar{\sigma}^a = (i,i)$, while in a 4-D CFT the usual choice is $\sigma^a = (\mathbb{1}; \vec{\sigma})$ and $\bar{\sigma}^a = (\mathbb{1}; -\vec{\sigma})$.

Given these ingredients we can build a set of bulk Dirac matrices $\Gamma^A$ by setting
\begin{equation}
\label{AdSoddDirac}
\Gamma^a = \gamma^a, \quad \Gamma^Z = -i\gamma.
\end{equation}
That is, the chirality operator in the CFT becomes ($i$ times) the radial Dirac matrix.  By construction these matrices satisfy $\lbrace \Gamma^A, \Gamma^B \rbrace = - 2 \eta^{AB} \mathbb{1}$.

Now let's solve the Dirac equation.  With the Dirac operator (\ref{DiracOp}) and the Dirac matrices (\ref{AdSoddDirac}) the Dirac equation $(i \Gamma^A e_A^M D_M - m)\Psi = 0$ becomes
\begin{equation}
\label{AdSoddDiracEqn}
\left(\begin{array}{cc}
-Z \partial_Z + {d \over 2} - m & i Z \sigma^a \partial_a \\
i Z \bar{\sigma}^a \partial_a & Z \partial_Z - {d \over 2} - m
\end{array}\right)
\left(\begin{array}{c} \Psi_- \\ \Psi_+ \end{array}\right) = 0.
\end{equation}
Note that on the boundary the subscripts $\psi_-$, $\psi_+$ indicate chirality, while in the bulk by (\ref{AdSoddDirac}) the subscripts $\Psi_-$, $\Psi_+$ indicate the $\Gamma^Z$ eigenvalue.
We adopt the general solution\footnote{
$J_\nu(x)$ are Bessel functions of the first kind. In checking that the Dirac equation is satisfied, a useful identity is
\begin{equation} x\partial_x J_\nu(x) = \mp \nu J_\nu(x) \pm x J_{\nu \mp 1}(x).\end{equation} For more details, see Appendix \ref{appendix:BF}.}
\begin{equation}
\label{AdSoddModeExpansion}
\Psi(T,\textbf{X},Z) =  \sum_s \int\limits_{\vert \omega \vert > \vert \textbf{k} \vert} \!\!\!d\omega d^{d-1}k \, a_{s \omega \textbf{k}} e^{-i \omega T} e^{i \textbf{k} \cdot \textbf{X}} Z^{d+1 \over 2}
\left(\begin{array}{c}
- \sigma^a p_a \epsilon_{s \omega \textbf{k}} \, J_{m+1/2}\left(\sqrt{\omega^2 - \vert \textbf{k} \vert^2} \, Z\right) \\ %[5pt]
\sqrt{\omega^2 - \vert \textbf{k} \vert^2} \, \epsilon_{s \omega {\bf k}} \, J_{m-1/2}\left(\sqrt{\omega^2 - \vert \textbf{k} \vert^2} \, Z\right)
\end{array}\right).
\end{equation}
Here $\sum_s$ is a sum over polarizations, $\epsilon_{s \omega k}$ are a complete set of positive-chirality boundary spinors, $a_{s \omega k}$ are the corresponding mode amplitudes,
and $\sigma^a p_a = - \sigma^0 \omega + \vec{\sigma} \cdot {\bf k}$.

A few comments on this mode expansion:
\begin{itemize}
\item
The mode expansion (\ref{AdSoddModeExpansion}) is valid provided $m$ is large enough for the field to be normalizable as $Z \rightarrow 0$.  This requirement is studied in Appendix \ref{appendix:BF}, where we show that the mode expansion in terms of $J$-type
Bessel functions is normalizable provided $m > - {1 \over 2}$.
\item
We also want fields that are well-behaved at the Poincar\'e horizon.  For this reason we have restricted $\vert \omega \vert > \vert {\bf k} \vert$ in
(\ref{AdSoddModeExpansion}), so that the arguments of the Bessel functions are real and the modes oscillate rather than grow exponentially as $Z \rightarrow \infty$.
\item
There's an analogous mode expansion in terms of $Y$-type Bessel functions which is normalizable provided $m < + {1 \over 2}$.  We will return to consider this alternate branch of solutions in Section \ref{sect:AdSoddAlternate}.
\end{itemize}

%%%%%%%%%%%%%%%%%%%%%%%%
\subsubsection{AdS${}_{\rm odd}$ extrapolate dictionary\label{sect:AdSoddExtrapolate}}
%%%%%%%%%%%%%%%%%%%%%%%%
To understand how the extrapolate dictionary works -- that is, to see how a bulk fermion turns into a CFT operator as $Z \rightarrow 0$ -- we study the near-boundary behavior
of the mode sum (\ref{AdSoddModeExpansion}).  As $Z \rightarrow 0$ we can use $J_\nu(x) \sim x^\nu$ to find that
\begin{equation}
\label{AdSoddextrapolate}
\Psi(T,{\bf X},Z) \stackrel{Z \rightarrow 0}{\sim} Z^{m+{d \over 2}} \left(\begin{array}{c} 0 \\ \psi_{+}(T,{\bf X}) \end{array}\right) + {\rm (subleading)}.
\end{equation}
We propose that $\psi_{+}$ should be identified with a CFT operator.  There are a few pieces of evidence in favor of this identification.
\begin{itemize}
\item
By construction $\psi_{+}$ is a positive-chirality spinor under the $SO(d-1,1)$ Lorentz group of the CFT.
\item
$\psi_{+}$ has a definite scaling dimension $\Delta = m + {d \over 2}$.  To see this note that
the AdS metric (\ref{Poincare}) is invariant under rescaling $(T,{\bf X},Z) \rightarrow \lambda (T,{\bf X},Z)$.  The Dirac action (\ref{action}) respects this symmetry, with $\Psi$ transforming
with weight zero under the rescaling.  On the boundary this transformation acts as a dilation, and from (\ref{AdSoddextrapolate}) we see that
$\psi$ transforms with weight $m+{d \over 2}$ under a dilation, $\psi \rightarrow \lambda^{-(m+{d \over 2})} \psi$.
\end{itemize}
This suggests that a Dirac fermion in the bulk extrapolates to a chiral fermion in the CFT.  That is,
we can identify $\psi_{+}$ with a positive-chirality spin-$1/2$ primary field of dimension $\Delta = m + {d \over 2}$ in the CFT.  

%%%%%%%%%%%%%%%%%%%%%%%%%
\subsubsection{AdS${}_{\rm odd}$ smearing functions\label{sect:AdSoddSmear}}
%%%%%%%%%%%%%%%%%%%%%%%%%
Our goal now is to recover the bulk field $\Psi(T,{\bf X},Z)$ from its boundary behavior $\psi_{+}(t,{\bf x})$.  This turns out to be relatively straightforward with the help of some calculations from \cite{Hamilton:2006fh}.

Our ingredients are the bulk mode expansion (\ref{AdSoddModeExpansion}) for the field $\Psi = \left({\Psi_- \atop \Psi_+}\right)$ and the definition (\ref{AdSoddextrapolate}) of the boundary operator $\psi_{+}$.  We start with the expansion for $\Psi_+$.  Denoting $\nu_\pm = m \pm \frac{1}{2}$ it reads
\begin{equation}
\label{AdSoddpsiplusmode}
\Psi_+(T,\textbf{X},Z) = \sum_s \int\limits_{\vert \omega \vert > \vert {\bf k} \vert} \!\!\!d\omega d^{d-1}k \, \epsilon_{s \omega {\bf k}} \, a_{s \omega {\bf k}} e^{-i \omega T} e^{i {\bf k} \cdot {\bf X}} Z^{d+1 \over 2}
\sqrt{\omega^2 - \vert {\bf k} \vert^2} \, J_{\nu_-}\big(\sqrt{\omega^2 - \vert {\bf k} \vert^2} \, Z\big).
\end{equation}
By sending $Z \rightarrow 0$ and stripping off a factor $Z^{m+{d \over 2}}$ we obtain an expression for $\psi_{+}$.
\begin{equation}
\psi_{+}(t,{\bf x}) = \sum_s \int\limits_{\vert \omega \vert > \vert {\bf k} \vert} \!\!\!d\omega d^{d-1}k \, \epsilon_{s \omega {\bf k}} \, a_{s \omega {\bf k}} e^{-i \omega t} e^{i {\bf k} \cdot {\bf x}}
{(\omega^2 - \vert {\bf k} \vert^2)^{\nu_+/2} \over
2^{\nu_-} \Gamma({\nu_- + 1})}.
\end{equation}
A Fourier transform gives
\begin{equation}
\label{AdSodda}
\sum_s \epsilon_{s \omega {\bf k}} \, a_{s \omega {\bf k}} = \int {dt d^{d-1}x \over (2\pi)^d} e^{i \omega t} e^{-i {\bf k} \cdot {\bf x}} {2^{\nu_-} \Gamma(\nu_- + 1) \over (\omega^2 - \vert {\bf k} \vert^2)^{\nu_+ / 2}} \psi_{+}(t,{\bf x}).
\end{equation}
Substituting this back in the expression (\ref{AdSoddpsiplusmode}) for $\psi_+$ we find that
\begin{equation}
\Psi_+(T,{\bf X},Z) = \int dt d^{d-1}x \, K_+(T,{\bf X},Z \vert t,{\bf x}) \psi_{+}(t,{\bf x}),
\end{equation}
where
\begin{equation}
\label{AdSoddKplus}
K_+ = Z^{1/2} \int\limits_{\vert \omega \vert > \vert {\bf k} \vert} \!\!\! {d\omega d^{d-1}k \over (2\pi)^d} \, e^{-i \omega (T-t)} e^{i {\bf k} \cdot ({\bf X} - {\bf x})}
{2^{\nu_-} \Gamma(\nu_- + 1) \over (\omega^2 - \vert {\bf k} \vert^2)^{\nu_-/2}} \, Z^{d/2} J_{\nu_-}\big(\sqrt{\omega^2 - \vert {\bf k} \vert^2} \, Z\big).
\end{equation}
Aside from the factor of $Z^{1/2}$ in front of (\ref{AdSoddKplus}), the same integrals appear in (9), (10) of \cite{Hamilton:2006fh}.\footnote{Set $\nu = \nu_-$, $\Delta = \nu_- + {d \over 2}$ in \cite{Hamilton:2006fh}.}  So from (21) in that paper we can read off
\begin{equation}
\label{AdSoddpsiplus}
\Psi_+(T,{\bf X},Z) = {Z^{1/2} \Gamma\big(m + {1 \over 2}\big) \over \pi^{d/2} \Gamma\big(m - {d \over 2} + {1 \over 2}\big)} \!\!\!\!\!\int\limits_{\,\,\,\,\,t^2 + \vert {\bf y} \vert^2 < Z^2} \!\!\!\!\!\!\!\!\!dtd^{d-1}y \left({Z^2 - t^2 - \vert {\bf y} \vert^2 \over Z}\right)^{m - {d \over 2} - {1 \over 2}} \psi_{+}(T + t,{\bf X} + i{\bf y}).
\end{equation}

It remains to obtain an expression for $\Psi_-$.  By following the same procedure, of substituting the expression (\ref{AdSodda}) for $\sum_s \epsilon_{s \omega {\bf k}} a_{s \omega {\bf k}}$ back into the mode expansion for
$\Psi_-$, we find that
\begin{equation}
\Psi_-(T,{\bf X},Z) = \int dt d^{d-1}x \, K_-(T,{\bf X},Z \vert t,{\bf x}) \psi_{+}(t,{\bf x}),
\end{equation}
where
\begin{equation}
K_- = {Z^{1/2} i \sigma^a \partial_a \over 2m+1}  \int\limits_{\vert \omega \vert > \vert {\bf k} \vert} \!\!\! {d\omega d^{d-1}k \over (2\pi)^d} \, e^{-i \omega (T-t)} e^{i {\bf k} \cdot ({\bf X} - {\bf x})}
{2^{\nu_+} \Gamma(\nu_+ + 1) \over (\omega^2 - \vert {\bf k} \vert^2)^{\nu_+/2}} \, Z^{d/2} J_{\nu_+}\big(\sqrt{\omega^2 - \vert {\bf k} \vert^2} \, Z\big).
\end{equation}
We've pulled a factor out in front, including a derivative that brings down a factor of $i p_a$.  The integral itself appears in (10) of \cite{Hamilton:2006fh}\footnote{Now
we're setting $\nu = \nu_+$, $\Delta = \nu_+ + {d \over 2}$ in \cite{Hamilton:2006fh}.} and leads to
\begin{equation}
\label{AdSoddpsiminus}
\Psi_-(T,{\bf X},Z) = {Z^{1/2} \Gamma\big(m + {1 \over 2}\big) \over 2 \pi^{d/2} \Gamma\big(m - {d \over 2} + {3 \over 2}\big)} \!\!\!\!\!\int\limits_{\,\,\,\,\,t^2 + \vert {\bf y} \vert^2 < Z^2} \!\!\!\!\!\!\!\!\!dtd^{d-1}y \left({Z^2 - t^2 - \vert {\bf y} \vert^2 \over Z}\right)^{m - {d \over 2} + {1 \over 2}} i \sigma^a \partial_a\psi_{+}(T + t,{\bf X} + i{\bf y}).
\end{equation}

Together (\ref{AdSoddpsiplus}) and (\ref{AdSoddpsiminus}) provide an expression for the bulk fermion in terms of the CFT operator $\psi_{+}$.  Just to summarize we've found
\begin{align}
\begin{split}\label{AdSoddsmear}
\Psi_-(T,{\bf X},Z) &= {Z^{1/2} \Gamma\big(m + {1 \over 2}\big) \over 2 \pi^{d/2} \Gamma\big(m - {d \over 2} + {3 \over 2}\big)} \!\!\!\!\!\int\limits_{\,\,\,\,\,t^2 + \vert {\bf y} \vert^2 < Z^2} \!\!\!\!\!\!\!\!\!dt d^{d-1}y \left({Z^2 - t^2 - \vert {\bf y} \vert^2 \over Z}\right)^{m - {d \over 2} + {1 \over 2}} i \sigma^a \partial_a\psi_{+}(T + t,{\bf X} + i{\bf y}) \\
\Psi_+(T,{\bf X},Z) &= {Z^{1/2} \Gamma\big(m + {1 \over 2}\big) \over \pi^{d/2} \Gamma\big(m - {d \over 2} + {1 \over 2}\big)} \!\!\!\!\!\int\limits_{\,\,\,\,\,t^2 + \vert {\bf y} \vert^2 < Z^2} \!\!\!\!\!\!\!\!\!dt d^{d-1}y \left({Z^2 - t^2 - \vert {\bf y} \vert^2 \over Z}\right)^{m - {d \over 2} - {1 \over 2}} \psi_{+}(T + t,{\bf X} + i{\bf y}).
\end{split}
\end{align}
There are a few consistency checks we can perform.
\begin{itemize}
\item
As $Z \rightarrow 0$ our fermion fields have the requisite behavior (\ref{AdSoddextrapolate}).  To see this note that the region of integration becomes very small as $Z \rightarrow 0$, so one can bring
the CFT operators out of the integral and explicitly verify the $Z \rightarrow 0$ behavior.
\item
The $SO(d-1,1)$ Lorentz representations work out: on the boundary $\sigma^a \partial_a \psi_{+}$ transforms as a spinor with negative chirality.  This is clear from the form of the CFT Dirac matrices
(\ref{CFTevenDirac}) which map a spinor of one chirality to a spinor of the opposite chirality.
\end{itemize}

Finally, we discuss the validity of (\ref{AdSoddsmear}).  As shown in Appendix \ref{appendix:BF}, normalizeability of the modes (\ref{AdSoddModeExpansion}) requires $m > - {1 \over 2}$.  But for the kernels in (\ref{AdSoddsmear}) to be integrable against smooth test functions we must impose a stronger
condition, namely that $m > {d \over 2} - {1 \over 2}$.  This condition also avoids poles in the gamma functions.  If the condition $m > {d \over 2} - {1 \over 2}$ is violated we expect that the smearing functions (\ref{AdSoddsmear}) should be replaced with distributions, as noted for scalar fields in \cite{Kabat:2012hp,Morrison:2014jha,DelGrosso:2019gow}.  But we leave an exploration of this issue to future work.

%%%%%%%%%%%%%%%%%%%%%%%%
\subsubsection{AdS${}_{\rm odd}$ fields -- alternate branch\label{sect:AdSoddAlternate}}
%%%%%%%%%%%%%%%%%%%%%%%%
There is another branch of solutions to the Dirac equation (\ref{AdSoddDiracEqn}), obtained by replacing $J_\nu$ with $Y_\nu$ in (\ref{AdSoddModeExpansion}).  As discussed in Appendix \ref{appendix:BF}, this other branch of solutions is normalizable provided $m < + {1 \over 2}$.  We could derive the smearing functions for this other branch by repeating the steps in Section \ref{sect:AdSoddSmear}.  But it's simpler to make use of a parity symmetry that takes $m \rightarrow -m$ and exchanges the two branches of solutions.

Given a bulk spinor field $\Psi$ we define the parity-transformed field $\Psi^P$ by
\begin{equation}
\label{parity}
\Psi^P(T,X_1,\ldots,X_{d-1},Z) = P \Psi(T,X_1,\ldots,X_{d-1},Z) \equiv i \Gamma^1 \Psi(T,-X_1,X_2,\ldots,X_{d-1},Z).
\end{equation}
This parity transformation reflects the $X_1$ spatial coordinate.  We've defined $P$ so that it's a symmetry of both the bulk and boundary metrics; the factor of $i$ is present in the definition so that $P^2 = 1$.  By acting with $P$ on the Dirac equation (\ref{AdSoddDiracEqn})
one finds that $\Psi^P$ satisfies exactly the same equation but with the replacement $m \rightarrow -m$.  That is,
\begin{equation}
i\left(Z \Gamma^a \partial_a + Z \Gamma^Z \partial_Z - {d \over 2} \Gamma^Z\right) \Psi = m \Psi
\end{equation}
implies
\begin{equation}
i\left(Z \Gamma^a \partial_a + Z \Gamma^Z \partial_Z - {d \over 2} \Gamma^Z\right) \Psi^P = - m \Psi^P.
\end{equation}
Thus parity is explicitly broken by the mass term.\footnote{This is true in odd spacetime dimensions, where parity changes the sign of the mass.  In even spacetime dimensions one has a chirality matrix $\Gamma$ and one can combine $P$ with $\Psi \rightarrow \Gamma \Psi$
to obtain a modified parity operation that leaves the mass invariant.}

In practice this means we can use our previous result (\ref{AdSoddsmear}) with the replacement $m \rightarrow -m$ to obtain an expression for $\Psi^P$ in terms of its near-boundary behavior, a result which in turn can be used to obtain an expression for $\Psi$ itself.  In doing this the representation (\ref{CFTevenDirac}), (\ref{AdSoddDirac}) and the algebra (\ref{sigmaAlgebra}) is crucial.  We find that the near-boundary behavior is
\begin{equation}
\label{AdSoddextrapolateAlternate}
\Psi(T,{\bf X},Z) \stackrel{Z \rightarrow 0}{\sim} Z^{-m+{d \over 2}} \left(\begin{array}{c} \psi_{-}(T,{\bf X}) \\ 0 \end{array}\right) + {\rm (subleading)}
\end{equation}
and that the bulk field can be reconstructed as $\Psi = \left({\Psi_- \atop \Psi_+}\right)$ with
\begin{align}
\begin{split}\label{AdSoddsmearAlternate}
 \Psi_-(T,{\bf X},Z) = {Z^{1/2} \Gamma\big(-m + {1 \over 2}\big) \over \pi^{d/2} \Gamma\big(-m - {d \over 2} + {1 \over 2}\big)} \!\!\!\!\!\int\limits_{\,\,\,\,\,t^2 + \vert {\bf y} \vert^2 < Z^2} \!\!\!\!\!\!\!\!\!dt d^{d-1}y \left({Z^2 - t^2 - \vert {\bf y} \vert^2 \over Z}\right)^{-m - {d \over 2} - {1 \over 2}}
\psi_{-}(T + t,{\bf X} + i{\bf y})\\
 \hspace{-12mm} \Psi_+(T,{\bf X},Z) = {Z^{1/2} \Gamma\big(-m + {1 \over 2}\big) \over 2\pi^{d/2} \Gamma\big(-m - {d \over 2} + {3 \over 2}\big)} \!\!\!\!\!\int\limits_{\,\,\,\,\,t^2 + \vert {\bf y} \vert^2 < Z^2} \!\!\!\!\!\!\!\!\!dt d^{d-1}y \left({Z^2 - t^2 - \vert {\bf y} \vert^2 \over Z}\right)^{-m - {d \over 2} + {1 \over 2}}
\left(-i\bar{\sigma}^a \partial_a \right) \psi_{-}(T + t,{\bf X} + i{\bf y}).
\end{split}
\end{align}
From (\ref{AdSoddextrapolateAlternate}) we see that $\psi_{-}$ can be identified with a negative-chirality spinor in the CFT with dimension $\Delta = - m + {d \over 2}$.  From Appendix \ref{appendix:BF}, unitarity in the CFT (or normalizeability in the bulk) requires $m < + {1 \over 2}$,
but convergence of the integral (\ref{AdSoddsmearAlternate}) places a stronger condition $m < {1 \over 2} - {d \over 2}$.

%%%%%%%%%%%%%%%%%%%%%%%%
\subsection{AdS${}_{\rm even}$ / CFT${}_{\rm odd}$\label{sect:AdSeven}}
%%%%%%%%%%%%%%%%%%%%%%%%
We now consider the case where the spacetime dimension of the CFT $d$ is odd.  The basic steps are the same as before but some new features arise from the nature of the Dirac matrices.
In particular, bulk spinors now have twice as many components as boundary spinors ($2^{d+1 \over 2}$ in the bulk compared to $2^{d-1 \over 2}$ on the boundary).
Our treatment will be somewhat brief and emphasize the differences from the previous case.

In the CFT we have a set of $d$ Dirac matrices $\gamma^a$ satisfying $\lbrace \gamma^a, \gamma^b \rbrace = - 2 \eta^{ab} \mathbb{1}$.  We can build a set of bulk Dirac matrices by setting
\begin{align}\label{AdSevenDirac}
\Gamma^a = \left(\begin{array}{cc} 0 & \gamma^a \\ \gamma^a & 0 \end{array}\right), \quad
\Gamma^Z = \left(\begin{array}{cc} 0 & \mathbb{1} \\ \mathbb{-1} & 0 \end{array}\right).
\end{align}
In the bulk we also have a chirality operator
\begin{equation}
\Gamma = \left(\begin{array}{cc} -\mathbb{1} & 0 \\ 0 & \mathbb{1} \end{array}\right).
\end{equation}
It's straightforward to check that these matrices satisfy $\lbrace \Gamma^A, \Gamma^B \rbrace = - 2 \eta^{AB} \mathbb{1}$, $\lbrace \Gamma, \Gamma^A \rbrace = 0$.
We'll write a bulk Dirac spinor in terms of its chiral components as $\Psi = \left({\Psi_- \atop \Psi_+}\right)$.  Note that in AdS${}_{\rm even}$ the subscripts $\Psi_-$, $\Psi_+$ refer to
bulk chirality while in AdS${}_{\rm odd}$ they referred to boundary chirality or equivalently the bulk $\Gamma^Z$ eigenvalue.

Using the Dirac operator (\ref{DiracOp}) and the Dirac matrices (\ref{AdSevenDirac}), the Dirac equation $(i \Gamma^A e_A^M D_M - m)\Psi = 0$ becomes
\begin{equation}
\left(\begin{array}{cc}
-m & i\big(Z \gamma^a \partial_a + Z \partial_Z - {d \over 2}\big) \\
i\big(Z \gamma^a \partial_a - Z \partial_Z + {d \over 2}\big) & -m
\end{array}\right)
\left(\begin{array}{c} \Psi_- \\ \Psi_+ \end{array}\right) = 0.
\end{equation}
We take the general solution\footnote{Again in checking that the Dirac equation is satisfied, a useful Bessel identity is
\begin{equation} x\partial_x J_\nu(x) = \mp \nu J_\nu(x) \pm x J_{\nu \mp 1}(x)\end{equation}}
\begin{equation}
\label{AdSevenModeExpansion}
\Psi(T,\textbf{X},Z) =  \sum_s \int\limits_{\vert \omega \vert > \vert {\bf k} \vert} \!\!\!d\omega d^{d-1}k \, a_{s \omega {\bf k}} e^{-i \omega T} e^{i {\bf k} \cdot {\bf X}} Z^{d+1 \over 2}
\left(\begin{array}{c}
\epsilon_{s \omega {\bf k}} J_- + {i \gamma^a p_a \epsilon_{s \omega {\bf k}} \over \sqrt{\omega^2 - \vert {\bf k} \vert^2}} \, J_+ \\ %[5pt]
-i\epsilon_{s \omega {\bf k}} J_- - {\gamma^a p_a \epsilon_{s \omega {\bf k}} \over \sqrt{\omega^2 - \vert {\bf k} \vert^2}} \, J_+
\end{array}\right).
\end{equation}
Here $J_\pm = J_{m \pm 1/2}\big(\sqrt{\omega^2 - \vert {\bf k} \vert^2} \, Z\big)$, $\sum_s$ is a sum over polarizations, $\epsilon_{s \omega k}$ are a complete set of boundary spinors, $a_{s \omega k}$ are the corresponding mode amplitudes,
and $\gamma^a p_a = - \gamma^0 \omega + \vec{\gamma} \cdot {\bf k}$.

As before, normalizeability of the modes (\ref{AdSevenModeExpansion}) requires $m > - {1 \over 2}$.  But in AdS${}_{\rm even}$ one can flip the sign of the fermion mass by performing a chiral rotation.
This leads to an alternate branch of solutions which we discuss in Section \ref{sect:AdSevenAlternate}.

%%%%%%%%%%%%%%%%%%%%%%%%
\subsubsection{AdS${}_{\rm even}$ extrapolate dictionary\label{sect:AdSevenExtrapolate}}
%%%%%%%%%%%%%%%%%%%%%%%%
As $Z \rightarrow 0$ we can use $J_\nu(x) \sim x^\nu$ to find that
\begin{equation}
\label{AdSevenextrapolate}
\Psi(T,{\bf X},Z) \stackrel{Z \rightarrow 0}{\sim} Z^{m+{d \over 2}} \left(\begin{array}{c} \psi(T,{\bf X}) \\ -i \psi(T,{\bf X}) \end{array}\right) + {\rm (subleading)}.
\end{equation}
Given the representation of $\Gamma^Z$
in (\ref{AdSevenDirac}), this means that as $Z \rightarrow 0$ the bulk spinor approaches an eigenstate of $\Gamma^Z$ with eigenvalue $-i$
\begin{equation}
\Gamma^Z \Psi \sim -i \Psi \quad \text{as } Z \rightarrow 0.
\end{equation}
Up to a factor of $-i$, the two chiral components of the bulk spinor both asymptote to the same Dirac spinor $\psi$ on the boundary.
Following the same logic as before we propose that $\psi$ can be identified with a spin-$1/2$ primary field of dimension $\Delta = m + {d \over 2}$ in the CFT.

%%%%%%%%%%%%%%%%%%%%%%%%%
\subsubsection{AdS${}_{\rm even}$ smearing functions\label{sect:AdSevenSmear}}
%%%%%%%%%%%%%%%%%%%%%%%%%
To recover the bulk field $\Psi(T,\textbf{X},Z)$ from its boundary behavior $\psi(t,\textbf{x})$ we follow the same steps as before.  By sending $Z \rightarrow 0$ in the mode expansion (\ref{AdSevenModeExpansion})
and stripping off a factor $Z^{m + {d \over 2}}$ we obtain an expression for $\psi$.  A Fourier transform then gives ($\nu_\pm = m \pm {1 \over 2}$)
\begin{equation}
\label{AdSevena}
\sum_s \epsilon_{s \omega {\bf k}} \, a_{s \omega {\bf k}} = \int {dt d^{d-1}x \over (2\pi)^d} e^{i \omega t} e^{-i {\bf k} \cdot {\bf x}} {2^{\nu_-} \Gamma(\nu_- + 1) \over (\omega^2 - \vert {\bf k} \vert^2)^{\nu_- / 2}} \psi(t,{\bf x}).
\end{equation}
Substituting this back in the bulk mode expansion (\ref{AdSevenModeExpansion}) and bringing some trivial factors out in front\footnote{Factors of $Z^{1/2}$
and ${Z^{1/2} \gamma^a \partial_a \over 2m+1}$, in case you're curious.} we are left with the integrals (9), (10) of \cite{Hamilton:2006fh}, and from (21) in that paper we can read off an expression for the
bulk fermion.  To save writing it's convenient to define two combinations that essentially already appeared in (\ref{AdSoddsmear}), namely
\begin{align}
\begin{split}\label{psi1psi2}
\Psi_1(T,{\bf X},Z) &= {Z^{1/2} \Gamma\big(m + {1 \over 2}\big) \over \pi^{d/2} \Gamma\big(m - {d \over 2} + {1 \over 2}\big)} \!\!\!\!\!\int\limits_{\,\,\,\,\,t^2 + \vert {\bf y} \vert^2 < Z^2} \!\!\!\!\!\!\!\!\!dtd^{d-1}y \left({Z^2 - t^2 - \vert {\bf y} \vert^2 \over Z}\right)^{m - {d \over 2} - {1 \over 2}} \psi(T + t,{\bf X} + i{\bf y}) \\
\Psi_2(T,{\bf X},Z) &= {Z^{1/2} \Gamma\big(m + {1 \over 2}\big) \over 2 \pi^{d/2} \Gamma\big(m - {d \over 2} + {3 \over 2}\big)} \!\!\!\!\!\int\limits_{\,\,\,\,\,t^2 + \vert {\bf y} \vert^2 < Z^2} \!\!\!\!\!\!\!\!\!dtd^{d-1}y \left({Z^2 - t^2 - \vert {\bf y} \vert^2 \over Z}\right)^{m - {d \over 2} + {1 \over 2}} \gamma^a \partial_a \psi(T + t,{\bf X} + i{\bf y})\,.
\end{split}
\end{align}
In terms of these combinations the bulk Dirac spinor has chiral components given by
\begin{equation}
\label{AdSevensmear}
\Psi(T,{\bf X},Z) = \left(\begin{array}{c} \Psi_- \\ \Psi_+ \end{array}\right) = \left(\begin{array}{c} \Psi_1 + \Psi_2 \\ -i (\Psi_1 - \Psi_2) \end{array}\right).
\end{equation}
Just as before, for the kernels in (\ref{psi1psi2}) to be integrable against smooth test functions we must have $m > {d \over 2} - {1 \over 2}$.  If this condition is violated we expect that the smearing functions should be replaced with distributions, as noted for scalar fields in \cite{Kabat:2012hp,Morrison:2014jha,DelGrosso:2019gow}.

%%%%%%%%%%%%%%%%%%%%%%%%
\subsubsection{AdS${}_{\rm even}$ fields -- alternate branch\label{sect:AdSevenAlternate}}
%%%%%%%%%%%%%%%%%%%%%%%%
In AdS${}_{\rm even}$ an alternate branch of solutions can be obtained by replacing $J_\nu$ with $Y_\nu$ in (\ref{AdSevenModeExpansion}).  As shown in Appendix \ref{appendix:BF}, this alternate
branch is normalizable for $m < + {1 \over 2}$.  The simplest way to obtain smearing functions for this alternate branch is to use a symmetry which takes $m \rightarrow -m$ and exchanges the
two branches of solutions.  To this end we define a chirally-rotated bulk field
\begin{equation}
\Psi^\Gamma(T,{\bf X},Z) = \Gamma \Psi(T,{\bf X},Z).
\end{equation}
It's straightforward to check that $\Psi^\Gamma$ obeys a Dirac equation with the opposite sign for $m$.  So we can use (\ref{psi1psi2}), (\ref{AdSevensmear}) to obtain an expression for
$\Psi^\Gamma$ which in turn can be used to obtain an expression for $\Psi$.  We find that $\Psi$ has the near-boundary behavior
\begin{equation}
\label{AdSevenextrapolateAlternate}
\Psi(T,{\bf X},Z) \stackrel{Z \rightarrow 0}{\sim} Z^{-m+{d \over 2}} \left(\begin{array}{c} \psi(T,{\bf X}) \\ i \psi(T,{\bf X}) \end{array}\right) + {\rm (subleading)}.
\end{equation}
That is, as $Z \rightarrow 0$ the bulk spinor approaches an eigenstate of $\Gamma^Z$ with eigenvalue $+i$
\begin{equation}
\Gamma^Z \Psi \sim i \Psi \quad \text{as } Z \rightarrow 0.
\end{equation}
In terms of $\psi$ the bulk field can be expressed as
\begin{equation}
\label{AdSevensmearAlternate}
\Psi(T,{\bf X},Z) = \left(\begin{array}{c} \Psi_- \\ \Psi_+ \end{array}\right) = \left(\begin{array}{c} \Psi_1 + \Psi_2 \\ i (\Psi_1 - \Psi_2) \end{array}\right),
\end{equation}
where
\begin{align}
\begin{split}\label{psi1psi2Alternate}
\Psi_1(T,{\bf X},Z) &= {Z^{1/2} \Gamma\big(-m + {1 \over 2}\big) \over \pi^{d/2} \Gamma\big(-m - {d \over 2} + {1 \over 2}\big)} \!\!\!\!\!\int\limits_{\,\,\,\,\,t^2 + \vert {\bf y} \vert^2 < Z^2} \!\!\!\!\!\!\!\!\!dt d^{d-1}y \left({Z^2 - t^2 - \vert {\bf y} \vert^2 \over Z}\right)^{-m - {d \over 2} - {1 \over 2}} \psi(T + t,{\bf X} + i{\bf y}) \\
\Psi_2(T,{\bf X},Z) &= {Z^{1/2} \Gamma\big(-m + {1 \over 2}\big) \over 2 \pi^{d/2} \Gamma\big(-m - {d \over 2} + {3 \over 2}\big)} \!\!\!\!\!\int\limits_{\,\,\,\,\,t^2 + \vert {\bf y} \vert^2 < Z^2} \!\!\!\!\!\!\!\!\!dt d^{d-1}y \left({Z^2 - t^2 - \vert {\bf y} \vert^2 \over Z}\right)^{-m - {d \over 2} + {1 \over 2}} \gamma^a \partial_a\psi(T + t,{\bf X} + i{\bf y}).
\end{split}
\end{align}
From (\ref{AdSevenextrapolateAlternate}) we see that $\psi$ can be identified with a spinor of dimension $\Delta = - m + {d \over 2}$ in the CFT.  For this branch of solutions boundary unitarity (or bulk normalizeability) requires $m < + {1 \over 2}$ but convergence of the integral
(\ref{psi1psi2Alternate}) requires the stronger condition $m < {1 \over 2} - {d \over 2}$.

There is one aspect of this alternate branch which is a bit different from the way things worked in AdS${}_{\rm odd}$.  In AdS${}_{\rm odd}$ the two branches could be distinguished from the
CFT point of view by examining the chirality of the boundary fermion.  But in AdS${}_{\rm even}$ the CFT has no way to tell the two branches apart.  In AdS${}_{\rm even}$ it's best to think of the
transformation $\Psi \rightarrow \Gamma \Psi$ as a bulk field redefinition which relates the two branches in the bulk but acts trivially on the CFT.

%%%%%%%%%%%%%%%%%%%%%%%%%
\section{Recovering bulk correlators\label{sect:correlators}}
%%%%%%%%%%%%%%%%%%%%%%%%%
In this section we explicitly compute the bulk-boundary two-point function for Dirac spinors
\begin{align}\label{bulkbdy}
	\Braket{\Psi(T,\textbf{X},Z) \bar\psi(0)} = \int d t d^{d-1} x ~ K(T,\textbf{X},Z|t,\textbf{x}) 
	\Braket{ \psi_{+}(t, \textbf{x}) \bar\psi_{+}(0)}
\end{align}
in both AdS$_\text{even}$ and AdS$_\text{odd}$.
The necessary CFT boundary data is contained in the Dirac spinor boundary two-point function \cite{Ferrara:1972uq,Henningson:1998cd,Mueck:1998iz}
\begin{align}\label{cftdata}
	\Braket{ \psi(t,\textbf{x}) \bar\psi(0)} = \frac{\gamma^a \textbf{x}_a }{(-t^2 + |\textbf{x}|^2)^{m+\frac{d}{2}+\frac{1}{2}}}.
\end{align}
The application of the smearing functions to the boundary data involves $d$ dimensional integrals which can be evaluated using the same techniques as in \cite{Kabat:2011rz,Kabat:2012hp}.
The following identity will be useful
\begin{align}
\begin{split}\label{intident}
	&I(\alpha,\beta;T,\textbf{X},Z) \\
	=& \int\limits_{t^2 + |\textbf{y}|^2 < Z^2} d t d^{d-1} \textrm{y} ~ \left( \frac{Z^2 - t^2 - |\textbf{y}|^2}{Z}\right)^\alpha \frac{1}{\left[ -(T+t)^2+(\textbf{X}+ i \textbf{y})^2 \right]^\beta} \\
	=& \pi^{\frac{d}{2}} \frac{\Gamma(\alpha+1)}{\Gamma(\alpha+\frac{d}{2}+1)} \frac{Z^{\alpha+d}}{(-T^2+|\textbf{X}|^2)^\beta} \hyp{\beta,\beta-\frac{d}{2}+1; \alpha + \frac{d}{2} +1 ; - \frac{Z^2}{(-T^2 + |\textbf{X}|^2)}},
\end{split}
\end{align}
where $\hyp{a,b;c;x}$ is the ordinary hypergeometric function.

All correlators that we obtain are functions of the AdS invariant distance
\begin{align}\label{adsdist}
	\sigma(T,\textbf{X},Z|T',\textbf{X}',Z') = \frac{Z^2 + Z'^2 -(T-T')^2 + |\textbf{X}-\textbf{X}'|^2}{2 Z Z'},
\end{align}
as expected.

\subsection{AdS$_\text{odd}$}

We now compute the two-point function in CFT$_\text{even}$/AdS$_\text{odd}$ by applying the smearing functions \eqref{AdSoddsmear} to \eqref{cftdata}.
The boundary Dirac spinors in terms of their chiral components are $\psi = \begin{pmatrix}\psi_- \\ \psi_+ \end{pmatrix}$. Using the convention for the Dirac matrices \eqref{CFTevenDirac}, remembering that
\begin{align}
	\psi \bar{\phi} = \psi \phi^\dagger \gamma^0 = \begin{pmatrix}
	\psi_- \phi^\dagger_+ \bar{\sigma}^0 & \psi_- \phi^\dagger_- \sigma^0 \\
	\psi_+ \phi^\dagger_+ \bar{\sigma}^0 & \psi_+ \phi^\dagger_- \sigma^0
	\end{pmatrix},
\end{align}
we identify the lower-left quadrant as the relevant boundary two-point function of chiral spinors.
\begin{align}
\begin{split}\label{cftdataodd}
	\Braket{\psi_{+}(t,\textbf{x}) \psi^\dagger_{+}(0)} \bar{\sigma}^0 &= \frac{\bar{\sigma}^a \textbf{x}_a }{(-t^2 + |\textbf{x}|^2)^{m+\frac{d}{2}+\frac{1}{2}}} \\
	&= \mathcal{D} \frac{1}{(-t^2 + |\textbf{x}|^2)^{m+\frac{d}{2}-\frac{1}{2}}}.
\end{split}
\end{align}
The introduction of the differential operator
\begin{align}
	\mathcal{D} = -\frac{1}{2 m + d - 1} \bar{\sigma}^a \partial_a
\end{align}
allows us to apply \eqref{intident} in the upcoming integrals.

Using the constant $C_1 = \frac{\G{m+\frac{1}{2}}}{\pi^{d/2} \G{m-\frac{d}{2}+\frac{1}{2}}}$, we compute
\begin{align}
\begin{split}
	&\Braket{\Psi_{+}(T,\textbf{X},Z) \psi^\dagger_{+}(0)} \bar{\sigma}^0 \\
	=& C_1 Z^{1/2} \int\limits_{t^2 + |\textbf{y}|^2 < Z^2} d t d^{d-1} y ~
	   \left(
	\frac{Z^2 - t^2 - |\textbf{y}|^2}{Z}
	\right)^{m-\frac{d}{2}-\frac{1}{2}}
	\Braket{\psi_{+}(T+t,\textbf{X} + i \textbf{y}) \psi^\dagger_{+}(0)} \bar{\sigma}^0 \\
	=& C_1 Z^{1/2} \int\limits_{t^2 + |\textbf{y}|^2 < Z^2} d t d^{d-1} y ~ 	\left(
		\frac{Z^2 - t^2 - |\textbf{y}|^2}{Z}
		\right)^{m-\frac{d}{2}-\frac{1}{2}}
		\mathcal{D} \frac{1}{(-(T+t)^2 - |\textbf{X} + i \textbf{y}|^2)^{m+\frac{d}{2}-\frac{1}{2}}} \\
	=& C_1 Z^{1/2}
	\mathcal{D} I\left(m-\frac{d}{2}-\frac{1}{2},m+\frac{d}{2}-\frac{1}{2};T,\textbf{X},Z\right).
\end{split}
\end{align}
After acting with the differential operator $\mathcal{D}$, we arrive at the bulk-boundary two-point function\footnote{$\hyp{\alpha,\beta;\beta;x} = (1-x)^{-\alpha}$.
}
\begin{align}\label{psip0p}
	\Braket{\Psi_{+}(T,\textbf{X},Z) \psi^\dagger_{+}(0)} \bar{\sigma}^0 = Z^{m+\frac{d}{2}} \frac{\bar{\sigma}^a \textbf{X}_a }{(-T^2 + |\textbf{X}|^2 + Z^2)^{m+\frac{d}{2}+\frac{1}{2}}}.
\end{align}

Using the constant $C_2 = \frac{\G{m+\frac{1}{2}}}{2 \pi^{d/2} \G{m-\frac{d}{2}+\frac{3}{2}}}$, we obtain the other component
\begin{align}
\begin{split}
	&\Braket{\Psi_{-}(T,\textbf{X},Z) \psi^\dagger_{+}(0)} \bar{\sigma}^0\\
	=& C_2 Z^{1/2} \int\limits_{t^2 + |\textbf{y}|^2 < Z^2} d t ~ d^{d-1} \textrm{y}~
	\left(
	\frac{Z^2 - t^2 - |\textbf{y}|^2}{Z}
	\right)^{m-\frac{d}{2}+\frac{1}{2}}
	i \sigma^a \partial_a \Braket{\psi_{+}(T+t,\textbf{X} + i \textbf{y}) \psi^\dagger_{+}(0)}\bar{\sigma}^0 \\
	=& C_2 Z^{1/2} i \sigma^a \partial_a \mathcal{D} I\left(m-\frac{d}{2}+\frac{1}{2},m+\frac{d}{2}-\frac{1}{2}; T, \textbf{X}, Z\right).
\end{split}
\end{align}
The combination of differential operators $\sigma^a \partial_a \mathcal{D}$ is proportional to
\begin{align}
\begin{split}
	\sigma^a \partial_a \bar{\sigma}^b \partial_b &= \frac{1}{2} (\sigma^a \bar{\sigma}^b + \sigma^b \bar{\sigma}^a) \partial_a \partial_b \\
	&= -\eta^{m n} \partial_m \partial_n,
\end{split}
\end{align}
where we used \eqref{sigmaAlgebra}.
After acting with the differential operators,\footnote{$\frac{d}{d x} \hyp{\alpha,\beta;\gamma;x} = \frac{\alpha \beta}{\gamma}\hyp{\alpha+1,\beta+1;\gamma+1;x}$} we arrive at 
\begin{align}\label{psim0p}
	\Braket{\Psi_{-}(T,\textbf{X},Z) \psi^\dagger_{+}(0)} \bar{\sigma}^0 = i Z^{m+\frac{d}{2}+1} \frac{1}{(-T^2 + |\textbf{X}|^2 +Z^2)^{m+\frac{d}{2}+\frac{1}{2}}}.
\end{align}

Equations \eqref{psip0p} and \eqref{psim0p} are the only non-zero components of the bulk-boundary two-point function \eqref{bulkbdy}. Using the projection operator $P_\pm = \frac{1}{2}(\mathbb{1} \pm \gamma)$, where $\gamma$ is the chirality matrix \eqref{gammachiral}, we can summarize the results as
\begin{align}
\label{covariant}
\begin{split}
    \Braket{\Psi(T,\textbf{X},Z) \bar\psi(0)} &= Z^{m+\frac{d}{2}} \frac{\gamma^a \textbf{X}_a - i \gamma Z}{(-T^2 + |\textbf{X}|^2 + Z^2)^{m+\frac{d}{2}+\frac{1}{2}}} P_-\\
    &= Z^{m+\frac{d}{2}} \frac{\Gamma^A \textbf{X}_A P_-}{(-T^2 + |\textbf{X}|^2 + Z^2)^{m+\frac{d}{2}+\frac{1}{2}}}.
\end{split}
\end{align}
This computation confirms that the application of our smearing functions correctly reproduces the known results of \cite{Henningson:1998cd, Mueck:1998iz, Kawano:1999au}, which have been obtained through different methods.

We now check that we recover the boundary data from the bulk-boundary two-point functions in the boundary limit.
Starting with \eqref{psip0p}, we have
\begin{align}
    \lim_{Z \to 0} Z^{-m-\frac{d}{2}} \Braket{\psi_{+}(T,\textbf{X},Z) \psi^\dagger_{+}(0)} \bar{\sigma}^0
    = \frac{\bar{\sigma}^a \textbf{X}_a }{(-T^2 + |\textbf{X}|^2)^{m+\frac{d}{2}+\frac{1}{2}}},
\end{align}
which correctly reproduces \eqref{cftdataodd}. The boundary limit of \eqref{psim0p} vanishes due to the extra factor of $Z$.

\subsection{AdS$_\text{even}$}

The computations of two-point functions in AdS$_\text{even}$/CFT$_\text{odd}$ is analogous to the previous case. Using our conventions \eqref{AdSevenDirac}, we simply identify $\sigma$ and $\bar{\sigma}$ with $\gamma$, and the CFT operator is now the entire boundary spinor $\psi$.
The application of the smearing functions \eqref{psi1psi2} results in essentially the same integrals as above, and we obtain
\begin{align}
	\Braket{\Psi_1(T,\textbf{X},Z) \psi^\dagger(0)} \gamma^0 &= Z^{m+\frac{d}{2}} \frac{\gamma^a \textbf{X}_a}{(- T^2 + |\textbf{X}|^2 + Z^2)^{m+\frac{d}{2}+\frac{1}{2}}} \\
	\Braket{\Psi_2(T,\textbf{X},Z) \psi^\dagger(0)} \gamma^0 &= Z^{m+\frac{d}{2}+1} \frac{ 1}{(- T^2 + |\textbf{X}|^2 + Z^2)^{m+\frac{d}{2}+\frac{1}{2}}}.
\end{align}
The Dirac components can be given with \eqref{AdSevensmear}.  To present the result we define an enlarged boundary field with the same number of components as the bulk spinor.
\begin{equation}
\Psi_{\rm bdy} = \left({\psi \atop -i \psi}\right)
\end{equation}
Then the bulk-boundary correlator in AdS${}_{\rm even}$ can be presented in a form similar to (\ref{covariant}),
\begin{equation}
\Braket{\Psi(T,\textbf{X},Z) \bar\Psi_{\rm bdy}(0)} = Z^{m+\frac{d}{2}} \frac{2\,\Gamma^A \textbf{X}_A P_-}{(-T^2 + |\textbf{X}|^2 + Z^2)^{m+\frac{d}{2}+\frac{1}{2}}}
\end{equation}
where $P_- = \frac{1}{2}\big(\mathbb{1} - i \Gamma^Z\big)$ projects onto a definite $\Gamma^Z$ eigenvalue.

%%%%%%%%%%%%%%%%%%%%%%%%%
\section{CFT modular flow as a bulk Lie derivative\label{sect:modular}}
%%%%%%%%%%%%%%%%%%%%%%%%%
In this section we compute the action of the CFT modular Hamiltonian on the CFT representation of bulk fermion operators.  We compare the results to the action of a bulk Lie derivative along a
Killing vector on bulk spinors.  We find perfect agreement.

Before we start, let us state a few results that we will need later. The CFT modular Hamiltonian in the vacuum state, for a spherical region of radius $R$ centered around the origin, is given by
\begin{equation}
\frac{1}{2\pi}H_{\rm mod}=\frac{1}{2R}(Q_{0}-R^2P_{0}),
\end{equation}
where $Q_{0}$ generates a special conformal transformation in the time direction and $P_0$ generates a translation in the time direction.
The action of the CFT modular Hamiltonian on a CFT primary Dirac fermion is given by
\begin{equation}
\frac{1}{2\pi i}[H_{\rm mod},\psi(t,{\bf x})]=\frac{1}{2R}\left (({\bf x}^2+t^2-R^2)\partial_{t}+2tx^{i}\partial_{i}+2t\Delta-x^{i}\gamma^{0}\gamma^{i}\right ) \psi(t,{\bf x}),
\label{hmodfermi}
\end{equation}
where $\gamma^{a}$ are the Dirac matrices in the CFT and $\Delta$ is the conformal dimension of the primary fermion.
We define the smearing function
\begin{equation}
K_{\Delta,d}(Z,{\bf X},T|t,{\bf x}) =\Theta \left(\frac{Z^2+({\bf X}-{\bf x})^2-(T-t)^2}{Z}\right) \left(\frac{Z^2+({\bf X}-{\bf x})^2-(T-t)^2}{Z}\right)^{\Delta-d},
\end{equation}
where $\Theta(x)$ is the step function.
We state a result from \cite{Kabat:2017mun,Kabat:2018smf}. For any operator ${\cal O} (t,{\bf x})$ in the CFT,  with ${\bf x}={\bf X}+i{\bf y}$,
\begin{eqnarray}
&&\frac{1}{2R}\int dt d{\bf y}  K_{\Delta,d}(Z,{\bf X},T|t,{\bf x})
\left(({\bf x}^2+t^2-R^2)\partial_{t}+2{\bf x}t\partial_{{\bf x}}+2\Delta t \right){\cal O} (t,{\bf x})\nonumber\\
&&=\xi^{M}\partial_{M}
\int dt d{\bf y} K_{\Delta,d}(Z,{\bf X},T|t,{\bf x}){\cal O} (t,{\bf x}),
\label{scalarresult}
\end{eqnarray}
where 
\begin{equation}
\left(\xi^{T},\xi^{{\bf X}},\xi^{Z}\right)=\left(\frac{1}{2R}(Z^2+{\bf X}^2+T^2-R^2),\frac{T{\bf X}}{R},\frac{TZ}{R}\right).
\end{equation}
This can be generalized to a spherical region centered around ${\bf X}_{0}$ by replacing ${\bf X} \rightarrow {\bf X}-{\bf X}_{0}$ in the above result.

%%%%%%%%%%%%%%%%%%%%%%%%%%%%%%%
\subsection{AdS${}_{\rm odd}$ / CFT${}_{\rm even}$}
%%%%%%%%%%%%%%%%%%%%%%%%%%%%%%%
The CFT Dirac matrices are given by
\begin{equation}
\label{CFTevenDirac2}
\gamma^a = \left(\begin{array}{cc} 0 & \sigma^a \\ \bar{\sigma}^a & 0 \end{array}\right),
\end{equation}
where the matrices $\sigma^a$, $\bar{\sigma}^a$ satisfy
\begin{align}
\begin{split}\label{sigmaAlgebra2}
\sigma^a \bar{\sigma}^b + \sigma^b \bar{\sigma}^a &= - 2 \eta^{ab} \mathbb{1} \\
\bar{\sigma}^a \sigma^b + \bar{\sigma}^b \sigma^a &= -2 \eta^{ab} \mathbb{1}.
\end{split}
\end{align}
The action of the modular Hamiltonian on a primary chiral fermion is then given by
\begin{equation}
\frac{1}{2\pi i}[H_{\rm mod},\psi_+(t,{\bf x})]=\frac{1}{2R}\left (({\bf x}^2+t^2-R^2)\partial_{t}+2tx^{i}\partial_{i}+2t\left(\Delta-\frac{1}{2}\right) +(t-x^{i}\bar{\sigma}^{0}\sigma^{i})\right ) \psi_+(t,{\bf x}).
\label{hmodfermieven}
\end{equation}

From Section \ref{sect:AdSoddSmear}, the representation of a Dirac fermion in the bulk of AdS${}_{d+1}$ of mass $m=\Delta-\frac{d}{2}$, using ${\bf x}={\bf X}+i{\bf y}$, is given by
\begin{eqnarray}
\Psi_+(T, {\bf X},Z) &=& {Z^{1/2} \Gamma\big(\Delta-\frac{d}{2}+ {1 \over 2}\big) \over \pi^{d/2} \Gamma\big(\Delta - d+ {1 \over 2}\big)} \int dtd^{d-1}y K_{\Delta-\frac{1}{2},d}(Z,{\bf X},T|t,{\bf x})\psi_{+}(t ,{\bf x})\\
\Psi_-(T,{\bf X},Z) &=& {Z^{1/2} \Gamma\big(\Delta-\frac{d}{2} + {1 \over 2}\big) \over 2 \pi^{d/2} \Gamma\big(\Delta - d + {3 \over 2}\big)} \int dtd^{d-1}y K_{\Delta+\frac{1}{2},d}(Z,{\bf X},T|t,{\bf x}) i \sigma^a \partial_a\psi_{+}(t ,{\bf x}).
\label{AdSoddsmear2}
\end{eqnarray}

We now compute
\begin{align}
\begin{split}
\frac{1}{2\pi i}&[H_{\rm mod},\Psi_{+}(Z,{\bf X},T)]=\frac{1}{2R} {Z^{1/2} \Gamma\big(\Delta-\frac{d}{2}+ {1 \over 2}\big) \over \pi^{d/2} \Gamma\big(\Delta - d+ {1 \over 2}\big)} \int dtd^{d-1}y K_{\Delta-\frac{1}{2},d}(Z,{\bf X},T|t,{\bf x}) \\
&\times  \left ( \left\{({\bf x}^2+t^2-R^2)\partial_{t}+2tx^{i}\partial_{i}+2t\left(\Delta-\frac{1}{2}\right)\right\} \psi_+(t,{\bf x}) +(t-x^{i}\bar{\sigma}^{0}\sigma^{i}) \psi_+(t,{\bf x})\right ).
\end{split}
\end{align}
The first factor in curly brackets in the second line has the form (\ref{scalarresult}), thus it gives
\begin{equation}
Z^{1/2}\xi^{M}\partial_{M} (Z^{-1/2} \Psi_{+}(Z,{\bf X},T))=\xi^{M}\partial_{M} \Psi_{+}-\frac{T}{2R}\Psi_{+}.
\end{equation}
The second factor in the second line can be written as 
\begin{equation}
{Z^{1/2} \Gamma\big(\Delta-\frac{d}{2}+ {1 \over 2}\big) \over \pi^{d/2} \Gamma\big(\Delta - d+ {1 \over 2}\big)}  \int dtd^{d-1}y K_{\Delta-\frac{1}{2},d}(Z,{\bf X},T|t,{\bf x})((t-T)-(x^{i}-X^{i})\bar{\sigma}^{0}\sigma^{i}) \psi_+(t,{\bf x})+\frac{1}{2R}(T-X^{i}\bar{\sigma}^{0}\sigma^{i})\Psi_{+}.
\end{equation}
The first factor (involving the integral) can be written as
\begin{equation}
-\frac{Z}{2(\Delta-d+\frac{1}{2})}{Z^{1/2} \Gamma\big(\Delta-\frac{d}{2}+ {1 \over 2}\big) \over \pi^{d/2} \Gamma\big(\Delta - d+ {1 \over 2}\big)}  \int dtd^{d-1}y\left ((\partial_{t}+\bar{\sigma}^{0}\sigma^{i}\partial_{i})  K_{\Delta+\frac{1}{2},d}(Z,{\bf X},T|t,{\bf x})\right ) \psi_+(t,{\bf x}),
\end{equation}
which after integration by parts becomes
\begin{equation}
-i\frac{Z}{2R}\bar{\sigma}^{0}\Psi_{-}.
\end{equation}
Thus overall we find 
\begin{equation}
\frac{1}{2\pi i}[H_{\rm mod},\Psi_{+}]=\xi^{M}\partial_{M}\Psi_{+}-\frac{X^{i}}{2R}\bar{\sigma}^{0}\sigma^{i}\Psi_{+}-i\frac{Z}{2R}\bar{\sigma}^{0}\Psi_{-}.
\label{oddpsi+f}
\end{equation}

Let us now compute $\frac{1}{2\pi i}[H_{\rm mod},\Psi_{-}]$. For this, using equation (\ref{hmodfermi}), we compute
\begin{align}
\begin{split}
\frac{1}{2\pi i}[H_{\rm mod},\sigma^{a}\partial_{a} \psi_{+}]&=\frac{1}{2R}\left ( ({\bf x}^2+t^2-R^2)\partial_{t}+2tx^{i}\partial_{i}+2t\left(\Delta+\frac{1}{2}\right)\right ) \sigma^{a}\partial_{a} \psi_{+} \\
&+ \frac{1}{2R}\left (t \sigma^{a}\partial_{a} +(2\Delta-d+1)\sigma^{0}+\sigma^{j}x_{j}\partial_{t}-x^{i}\sigma^{0}\bar{\sigma}^{i}\sigma^{j}\partial_{j} \right )\psi_{+}
\end{split}
\end{align}
so
\begin{align}
\begin{split}\label{oddpsi-}
& \frac{1}{2\pi i}[H_{\rm mod},\Psi_{-}]= {Z^{1/2} \Gamma\big(\Delta-\frac{d}{2} + {1 \over 2}\big) \over 2 \pi^{d/2} \Gamma\big(\Delta - d + {3 \over 2}\big)}\int dtd^{d-1}y K_{\Delta+\frac{1}{2},d}(Z,{\bf X},T|t,{\bf x}) \\
& \Bigg\{ \frac{1}{2R}\left ( ({\bf x}^2+t^2-R^2)\partial_{t}+2tx^{i}\partial_{i}+2t\left(\Delta+\frac{1}{2}\right)\right ) i\sigma^{a}\partial_{a} \psi_{+} \\
+& \frac{1}{2R}\left (t \sigma^{a}\partial_{a} +(2\Delta-d+1)\sigma^{0}+\sigma^{j}x_{j}\partial_{t}-x^{i}\sigma^{0}\bar{\sigma}^{i}\sigma^{j}\partial_{j} \right )i\psi_{+}\Bigg\}.
\end{split}
\end{align}
The contribution of the second line of (\ref{oddpsi-}) has the form (\ref{scalarresult}), thus it contributes
\begin{equation}
Z^{1/2}\xi^{M}\partial_{M}(Z^{-1/2}\Psi_{-}).
\end{equation}
The third line can be written as 
\begin{align}
\begin{split}\label{oddpsi-2}
 &\frac{1}{2R} (T-X^{i}\sigma^{0}\bar{\sigma}^{i})i\sigma^{a}\partial_{a} \psi_{+} \\
 +& \frac{1}{2R}\left ((t-T) \sigma^{a}\partial_{a} +(2\Delta-d+1)\sigma^{0}+\sigma^{j}(x_{j}-X_{j})\partial_{t}-(x^{i}-X^{i})\sigma^{0}\bar{\sigma}^{i}\sigma^{j}\partial_{j} \right )i\psi_{+}.
\end{split}
\end{align}
The first line of (\ref{oddpsi-2}) results in a contribution of 
\begin{equation}
\frac{T}{2R}\Psi_{-}-\frac{X^i}{2R}\sigma^{0} \bar{\sigma}^{i} \Psi_{-}.
\end{equation}
The second line in  (\ref{oddpsi-2}) can be integrated by parts inside the integral of (\ref{oddpsi-}) and after some algebra it is seen to take the form
\begin{equation}
i\frac{Z}{2R}\sigma^{0}\Psi_{+}.
\end{equation}
Thus overall we find
\begin{equation}
\frac{1}{2 \pi i} [H_{\rm mod}, \Psi_{-}]=\xi^{M}\partial_{M}\Psi_{-}-\frac{X^i}{2R}\sigma^{0} \bar{\sigma}^{i} \Psi_{-}+i\frac{Z}{2R}\sigma^{0}\Psi_{+}.
\label{oddpsi-f}
\end{equation}

%%%%%%%%%%%%%%%%%%%%%%%%%%%%
\subsection{AdS${}_{\rm even}$ / CFT${}_{\rm odd}$}
%%%%%%%%%%%%%%%%%%%%%%%%%%%%
Here we start with a Dirac fermion $\psi(t,{\bf x})$  and the Dirac matrices $\gamma^{a}$ in the CFT.  The action of the modular Hamiltonian is
\begin{equation}
\frac{1}{2\pi i}[H_{\rm mod},\psi_{0}]=\frac{1}{2R}\left (({\bf x}^2+t^2-R^2)\partial_{t}+2tx^{i}\partial_{i}+2t\Delta-x^{i}\gamma^{0}\gamma^{i}\right ) \psi(t,{\bf x})
\end{equation}
from which we can compute
\begin{align}
\begin{split}
 & \frac{1}{2\pi i}[H_{\rm mod},\gamma^{a}\partial_{a}\psi_{0}]=\frac{1}{2R}\left (({\bf x}^2+t^2-R^2)\partial_{t}+2tx^{i}\partial_{i}+2t(\Delta+\frac{1}{2})\right ) \gamma^{a}\partial_{a}\psi(t,{\bf x}) \\
& \quad + \frac{1}{2R}\left( t\gamma^{a}\partial_{a}+(2\Delta-d+1)\gamma^{0}+\gamma^{i}x_{i}\partial_{t}+x_{i}\gamma^{i}\gamma^{0}\gamma^{j}\partial_{j}\right) \psi(t,{\bf x}).
\end{split}
\end{align}
Starting with the expressions for $\Psi_1$ and $\Psi_2$ in Section \ref{sect:AdSevenSmear}, with steps similar to the computation in AdS${}_{\rm odd}$, we find
\begin{equation}
\frac{1}{2\pi i}[H_{\rm mod},\Psi_{1}(T,{\bf X},Z)]=\xi^{M}\partial_{M}\Psi_{1}-\frac{X_{i}}{2R}\gamma^{0}\gamma^{i}\Psi_{1}+\frac{Z}{2R}\gamma^{0}\Psi_{2} 
\label{b4psi1}
\end{equation}
\begin{equation}
\frac{1}{2\pi i}[H_{\rm mod},\Psi_{2}(T,{\bf X},Z)]=\xi^{M}\partial_{M}\Psi_{2}-\frac{X_{i}}{2R}\gamma^{0}\gamma^{i}\Psi_{2}+\frac{Z}{2R}\gamma^{0}\Psi_{1} .
\label{b4psi2}
\end{equation}

%%%%%%%%%%%%%%%%%%%%%%%%%%%%
\subsection{Bulk spinor Lie derivative}
%%%%%%%%%%%%%%%%%%%%%%%%%%%%
Given  a bulk Killing vector field $\xi^{M}$, one can define a natural  bulk spinor Lie derivative
\begin{equation}
{\cal L}_{\xi}\Psi=\xi^{M}\nabla_{M}\Psi-\frac{1}{8}(\nabla_{M}\xi_{N}-\nabla_{N}\xi_{M})e^{M}_{A}e^{N}_{B}\Gamma^{A}\Gamma^{B} \Psi,
\end{equation}
where
\begin{equation}
\nabla_{M}=\partial_{M}-\frac{1}{8}\omega^{AB}_{M}[\Gamma_{A},\Gamma_{B}], \ \ \{\Gamma_{A},\Gamma_{B}\}=-2\eta_{AB}.
\end{equation}
The $(M,N)$ indices are raised and lowered with the bulk metric and the $(A,B)$ indices are raised and lowered with $\eta_{AB}$.
Our Killing field is given by
\begin{equation}
\left(\xi^{T},\xi^{{\bf X}},\xi^{Z}\right)=\left(\frac{1}{2R}\left(Z^2+{\bf X}^2+T^2-R^2\right),\frac{T{\bf X}}{R},\frac{TZ}{R}\right).
\end{equation}
Using all this, we find the bulk result
\begin{equation}
{\cal L}_{\xi}\Psi=\xi^{M}\partial_{M}\Psi-\frac{X^{j}}{2R}\Gamma^{0}\Gamma^{j}\Psi-\frac{Z}{2R}\Gamma^{0}\Gamma^{Z}\Psi.
\end{equation}

%%%%%%%%%%%%%%%%%%%%%%%%%%%%
\subsubsection{AdS${}_{\rm odd}$}
%%%%%%%%%%%%%%%%%%%%%%%%%%%%
For AdS${}_{\rm odd}$ we have 
\begin{equation}
\Gamma^{a}=
\begin{pmatrix}
0& \sigma_{a} \\
\bar{\sigma}_{a} & 0
\end{pmatrix}, \quad  
\Gamma^{Z}=i
\begin{pmatrix}
\mathbb{1}& 0 \\
0 & -\mathbb{1}
\end{pmatrix},
\end{equation}
thus we find 
\begin{eqnarray}
{\cal L}_{\xi}\Psi_{+}=\xi^{M}\partial_{M}\Psi_{+}-\frac{X^{j}}{2R}\bar{\sigma}^{0}\sigma_{j}\Psi_{+}-\frac{iZ}{2R}\bar{\sigma}^{0}\Psi_{-} \\
{\cal L}_{\xi}\Psi_{-}=\xi^{M}\partial_{M}\Psi_{-}-\frac{X^{j}}{2R}\sigma^{0}\bar{\sigma}_{j}\Psi_{-}+\frac{iZ}{2R}\sigma^{0}\Psi_{+}
\end{eqnarray}
which agrees with (\ref{oddpsi+f}) and (\ref{oddpsi-f}).

%%%%%%%%%%%%%%%%%%%%%%%%%%%%
\subsubsection{AdS${}_{\rm even}$}
%%%%%%%%%%%%%%%%%%%%%%%%%%%%
For AdS${}_{\rm even}$ we have 
\begin{equation}
\Gamma^{a}=
\begin{pmatrix}
0& \gamma^{a} \\
\gamma^{a} & 0
\end{pmatrix}, \quad
\Gamma^{Z}=
\begin{pmatrix}
0& \mathbb{1} \\
-\mathbb{1}& 0
\end{pmatrix},
\end{equation}
thus we find
\begin{eqnarray}
{\cal L}_{\xi}\Psi_{+}&=&\xi^{M}\partial_{M}\Psi_{+}-\frac{X^{j}}{2R}\gamma^{0}\gamma^{j}\Psi_{+}-\frac{Z}{2R}\gamma^{0}\Psi_{+} \\
{\cal L}_{\xi}\Psi_{-}&=& \xi^{M}\partial_{M}\Psi_{-}-\frac{X^{j}}{2R}\gamma^{0}\gamma^{j}\Psi_{-}+\frac{Z}{2R}\gamma^{0}\Psi_{-}
\end{eqnarray}
With  $\Psi_{1}=\frac{1}{2}(\Psi_{-}+i\Psi_{+})$ and $\Psi_{2}=\frac{1}{2}(\Psi_{-}-i\Psi_{+})$ we recover (\ref{b4psi1}) and (\ref{b4psi2}).

%%%%%%%%%%%%%%%%%%%%%%%%%
\section{Chirality and reality conditions\label{sect:conditions}}
%%%%%%%%%%%%%%%%%%%%%%%%%
So far we've considered Dirac spinors in AdS.  We've developed smearing functions appropriate to either the $J_\nu$ or $Y_\nu$ branch of solutions, finding that
\begin{itemize}
\item
A Dirac spinor in AdS${}_{\rm odd}$ is dual to a chiral spinor in CFT${}_{\rm even}$.
\item
A Dirac spinor in AdS${}_{\rm even}$ is dual to a Dirac spinor (which has half as many components) in CFT${}_{\rm odd}$.
\end{itemize}
But as reviewed in Appendix \ref{appendix:spinor}, depending on the number of dimensions it may be possible to impose additional conditions on the bulk spinor representation: that it be chiral, or real, or both.  And as reviewed in Appendix \ref{appendix:BF}, a fermion in AdS must satisfy one of a limited choice of boundary conditions.  Our goal in this section is to figure
out when a spinor projection in the bulk is compatible with AdS boundary conditions.  In cases where they're compatible we want to identify the corresponding projection condition
in the CFT.

We begin by summarizing our results.  The compatible spinor projections are as follows.
\begin{itemize}
\item
A Dirac spinor in AdS${}_3$ is dual to a chiral spinor on the boundary.  A reality condition can be imposed on both
sides of the duality, so a Majorana spinor in AdS${}_3$ is dual to a Majorana - Weyl spinor on the boundary.
\item
A Dirac spinor in AdS${}_4$ is dual to a Dirac spinor on the boundary.  A reality condition can be imposed on both sides,
so a Majorana spinor in AdS${}_4$ is dual to a Majorana spinor in CFT${}_3$.  In AdS${}_4$ Majorana and chiral spinors are
equivalent, so a chiral spinor in AdS${}_4$ can also be expressed in terms of a Majorana spinor in CFT${}_3$.
\item
A Dirac spinor in AdS${}_{10}$ is dual to a Dirac spinor on the boundary.  A reality condition can be imposed on both sides,
so a Majorana spinor in AdS${}_{10}$ is dual to a Majorana spinor in CFT${}_9$.
\end{itemize}
The list above exhausts the distinct possibilities, with the pattern of compatible projections repeating in dimension mod 8.

The argument leading to this list is as follows. As reviewed in Appendix \ref{appendix:BF}, fermions in AdS must satisfy one of two types of boundary conditions
\begin{align}
\label{AdSbcJ}
& \hbox{\rm J type:} \quad\, \Gamma^Z \Psi \doteq - i \Psi \quad \hbox{\rm as $Z \rightarrow 0$} \\
\label{AdSbcY}
& \hbox{\rm Y type:} \quad \Gamma^Z \Psi \doteq  i \Psi \quad \hbox{\rm as $Z \rightarrow 0$}.
\end{align}
(By $\doteq$ we mean the coefficient of the leading $Z \rightarrow 0$ behavior of the two sides is the same.)

First we consider chiral spinors in AdS${}_{\rm even}$.  Could a bulk field be an eigenstate of the chirality operator $\Gamma$?  Starting from $\Gamma^Z \Psi \doteq \pm i \Psi$ and applying $\Gamma$
to both sides gives $\Gamma^Z \Gamma \Psi \doteq \mp i \Gamma \Psi$.  That is, $\Psi$ and $\Gamma \Psi$ obey opposite boundary conditions, as we indeed saw in Section \ref{sect:AdSevenAlternate}.  So AdS boundary conditions are not compatible with having a bulk spinor of definite chirality.\footnote{Note that this argument only rules out a single chiral spinor in the bulk.  Multiple chiral spinors may be
possible with more complicated boundary conditions.  A simple example is a bulk Dirac fermion $\Psi_{\rm Dirac}$, which can be thought of as two chiral spinors with opposite chirality that are related by the boundary
condition imposed on $\Psi_{\rm Dirac}$.}

Next consider imposing a Majorana condition.  We start from
\begin{equation}
\Gamma^Z \Psi \doteq \pm i \Psi
\end{equation}
and apply charge conjugation $\Psi^C = X \Psi^*$ to both sides.  Here $X = D$ or $X = \tilde{D}$ as reviewed in Appendix \ref{appendix:spinor}.  This leads to
\begin{equation}
X (\Gamma^{Z})^* X^{-1} \, \Psi^C \doteq \mp i \Psi^C.
\end{equation}
So a Majorana condition $\Psi = \Psi^C$ is compatible with AdS boundary conditions if and only if $X (\Gamma^{Z})^* X^{-1} = - \Gamma^Z$. From Appendix \ref{appendix:spinor},
this only happens in AdS${}_{d+1}$ if the bulk spacetime dimension satisfies $d+1 = 2,3,4\,\, \hbox{\rm(mod 8)}$.  This leads to the possibilities for Majorana spinors given above.  A curious
fact, which appears unrelated: this is also the list of dimensions in which a Majorana mass term is possible, see Appendix \ref{appendix:spinor} for details.

Finally, there is one special case to consider.  A chiral spinor is not compatible with the AdS boundary conditions (\ref{AdSbcJ}), (\ref{AdSbcY}).  But in AdS${}_4$ a Majorana spinor
is allowed, and moreover in AdS${}_4$ chiral and Majorana spinors are equivalent as Lorentz representations.  So there must be a consistent boundary condition that describes a chiral
spinor in AdS${}_4$.  The appropriate boundary condition is\footnote{For a Majorana spinor the usual boundary condition $\Gamma^Z \Psi \doteq \pm i \Psi$ is equivalent to
$\Gamma^Z \Psi_{\rm Majorana} \doteq \pm i \Psi^C_{\rm Majorana}$.  Setting $\Psi_{\rm Majorana} = \Psi_{\rm chiral} + \Psi_{\rm chiral}^C$ and projecting onto one spinor chirality
gives (\ref{AdSchiralbc}).}
\begin{equation}
\label{AdSchiralbc}
\Gamma^Z \Psi_{\rm chiral} \doteq \pm i \Psi^C_{\rm chiral}.
\end{equation}
This boundary condition is only possible if $d + 1 = 4\,\, \hbox{\rm(mod 8)}$.\footnote{It relies on charge conjugation changing the chirality of a spinor, so that both sides of (\ref{AdSchiralbc})
have the same chirality.  This only happens in $d + 1 = 0,4\,\, \hbox{\rm(mod 8)}$.  But taking the charge conjugate of (\ref{AdSchiralbc}) and multiplying by $\Gamma^Z$ requires
$X (\Gamma^{Z})^* X^{-1} = - \Gamma^Z$, which rules out the possibility $d + 1 = 0\,\, \hbox{\rm(mod 8)}$.}  It accounts for the last case given above, of a chiral spinor in AdS${}_4$ being dual to a Majorana
spinor in CFT${}_3$.

%%%%%%%%%%%%%%%%%%%%%%%%%
\section{Conclusions\label{sect:conclusions}}
%%%%%%%%%%%%%%%%%%%%%%%%%
In this paper we developed an approach to reconstructing bulk fermions in AdS in terms of the CFT.
We obtained smearing functions that let us represent a free bulk fermion as an operator in the CFT,
calculated the bulk-boundary correlators,
showed that bulk fermions transform as expected under modular flow,
and saw that in some cases chirality and reality conditions can be imposed.
We conclude with a few directions for further development.

In this paper we only considered free fermions in the bulk.  To describe an interacting bulk fermion, presumably one has to add a tower of higher-dimension multi-trace operators.
This has been shown in considerable detail for scalar fields \cite{Kabat:2015swa}.
It would be interesting to identify the appropriate CFT operators and carry out the construction of interacting spin-$1/2$ fields.

In Section \ref{sect:modular}, we showed that the bulk fermions we have constructed transform as expected under boundary modular flow. It should be possible to reverse the logic and
use modular flow as a starting point for constructing bulk fermions, along the lines used in \cite{Kabat:2018smf}, to construct bulk massive vectors by considering intersecting RT surfaces.

%%%%%%%%%%%%%%%%%%%
\bigskip
\goodbreak
\centerline{\bf Acknowledgements}
\noindent
DK thanks the Columbia University Center for Theoretical Physics for hospitality during this work.
The work of VF is supported by the James Arthur Graduate Award.
The work of DK is supported by U.S.\ National Science Foundation grant PHY-1820734.
The work of GL is supported in part by the Israel Science Foundation under grant 447/17.

\appendix
%%%%%%%%%%%%%%%%%%%%%%%%%
\section{Conventions for spin matrices\label{appendix:spin}}
%%%%%%%%%%%%%%%%%%%%%%%%%
The covariant derivative in any representation can be written in terms of a set of spin matrices $\Sigma_{ab}$
\begin{equation}
D_\mu = \partial_\mu + {1 \over 2} \omega_\mu^{ab} \Sigma_{ab}.
\end{equation}
In the vector representation we took $D_\mu v^a = \partial_\mu v^a + \omega_{\mu b}^a v^b$ (this is the representation that was used in (\ref{torsion}) to determine the
spin connection).  This means that in the vector representation we're using
\begin{equation}
(\Sigma_{ab})^c{}_d = \delta^c_a \eta_{bd} - \delta^c_b \eta_{ad}.
\end{equation}
These matrices generate the Lorentz algebra (suppressing the matrix indices)
\begin{equation}
[\Sigma_{ab},\Sigma_{cd}] = - \eta_{ac} \Sigma_{bd} + \eta_{ad} \Sigma_{bc} + \eta_{bc} \Sigma_{ad} - \eta_{bd} \Sigma_{ac}.
\end{equation}
Given a set of Dirac matrices obeying $\lbrace \Gamma^a,\Gamma^b \rbrace = - 2 \eta^{ab}$ one can check that by setting
\begin{equation}
\label{SpinorSigma}
\Sigma_{ab} = - {1 \over 4} [\Gamma_a,\Gamma_b]
\end{equation}
we obtain exactly the same algebra.  This shows that (\ref{SpinorSigma}) are the appropriate spin matrices to use in the spinor representation and leads to the spinor covariant derivative (\ref{SpinorD}).
In comparing to the literature note that the minus sign appearing in the Clifford algebra is correlated with the minus sign appearing in the spin matrices.

%%%%%%%%%%%%%%%%%%%%%%%%%
\section{The mass spectrum\label{appendix:BF}}

In this section we derive the spectrum of free, massive fermions in the AdS$_{d+1}$/CFT$_d$ correspondence and derive the analog to the BF bound for fermions.
The case of free, massive scalar fields in AdS was first discussed in \cite{Breitenlohner:1982jf}. It was shown that the theory is unitary only if the mass obeys $m^2 > -\frac{d^2}{4}$. In addition, it was shown that in the window of $-\frac{d^2}{4} < m^2 < -\frac{d^2}{4}+1$ two AdS-invariant quantizations are possible, while for $m^2>-\frac{d^2}{4} + 1$ only a single, unique solution is admissible.
We will now derive the analog constraints for the fermionic case.

\subsection{The Dirac equation}
We conduct our analysis of fermions in the Poincar\'e patch of Lorentzian AdS$_{d+1}$. The equation of motion associated with the action \eqref{action} is the Dirac equation
\begin{align}
	\left(i \Gamma^A e_A^M D_M - m \right) \Psi = 0,
\end{align}
where the Dirac operator is given by
\begin{align}
	\Gamma^A e^M_A D_M &= Z \Gamma^A \partial_A  - {d \over 2} \Gamma^Z\\
	&= Z \Gamma^a \partial_a + Z \Gamma^Z \partial_Z - {d \over 2} \Gamma^Z.
\end{align}
The components of the spinor also satisfy the squared Dirac equation
\begin{align}
	\left(i \Gamma^A e_A^M D_M - m \right) \left(i \Gamma^A e_A^M D_M + m \right) \Psi = 0,
\end{align}
which can be written as the Klein-Gordon-like equation
\begin{align}\label{diracsquared}
	\left[
	\eta^{M N} \partial_M \partial_N - \frac{d}{Z} \partial_Z + \frac{1}{Z^2} \left(
	\frac{d^2}{4} + \frac{d}{2}- m^2 + i m \Gamma^Z
	\right)
	\right] \Psi = 0.
\end{align}
Using the ansatz
\begin{align}
	\Psi(T,\textbf{X},Z) = e^{-i \omega T} e^{i \textbf{k} \cdot \textbf{X}} \phi(Z),
\end{align}
we obtain the second order differential equation
\begin{align}\label{modkg}
	\phi'' -\frac{d}{Z} \phi' + \left( q^2 - \frac{M^2}{Z^2} \right)\phi = 0,
\end{align}
where we denoted $\gamma = i \Gamma^Z, M^2 = m^2 - \gamma m -\frac{d^2}{4}-\frac{d}{2}$ and $q^2 = \omega^2 - |\textbf{k}|^2$. It is implied that all terms except $\gamma$ are multiples of the unit matrix.
With a basis of eigenspinors  $\gamma \epsilon{\pm} = \pm \epsilon_{\pm}$, the solution to \eqref{modkg} is
\begin{align}\label{kgsol}
	\Psi(T,\textbf{X}, Z) = e^{- i \omega T} e^{i \textbf{k} \cdot \textbf{X}} Z^{\frac{d+1}{2}}
	\left[ a_1 \epsilon_+ J_{\nu_+}(q Z) + a_2 \epsilon_- J_{\nu_-}(q Z) \right],
\end{align}
where $J_\alpha(x)$ are Bessel functions of the first kind, $\nu_\pm = m \pm \frac{1}{2}$ and $a_i$ are constants.
There is a second branch of solutions in terms of Bessel functions of the second kind $Y_\alpha(x)$ with constants $b_i$
\begin{align}\label{kgsol2}
	\Psi(T,\textbf{X}, Z) = e^{- i \omega T} e^{i \textbf{k} \cdot \textbf{X}} Z^{\frac{d+1}{2}}
	\left[ b_1 \epsilon_+ Y_{\nu_+}(q Z) + b_2 \epsilon_- Y_{\nu_-}(q Z) \right].
\end{align}
Since the Dirac equation is of first order, the constants are not independent for each branch. By plugging the solutions of the Klein-Gordon equation into the Dirac equation, we find the mode expansion \eqref{AdSoddModeExpansion} and \eqref{AdSevenModeExpansion} for AdS$_\text{odd}$ and AdS$_\text{even}$, respectively.

\subsection{Boundary behavior and normalizability}

For $0 < x \ll \sqrt{\nu+1}$, the Bessel functions behave like
\begin{align}
    \begin{split}
        J_\nu(x) \sim x^\nu, \quad Y_\nu(x) \sim x^{-\nu}.
    \end{split}
\end{align}
The behavior near the boundary ($Z \to 0$) of \eqref{kgsol} and \eqref{kgsol2} is thus
\begin{align}
\begin{split}\label{spinorasym}
	\Psi(T,\textbf{X},Z) \to 
		&\epsilon_+ \left( Z^{m+\frac{d}{2}+1} \left[ \psi_{-}(T,\textbf{X}) + \mathcal{O}(Z^2)] + Z^{-m+\frac{d}{2}} [\chi_{-}(T,\textbf{X}) + \mathcal{O}(Z^2) \right] \right) \\
		+&\epsilon_- \left( Z^{m+\frac{d}{2}} \left[ \psi_{+}(T,\textbf{X}) + \mathcal{O}(Z^2)] + Z^{-m+\frac{d}{2}+1} [\chi_{+}(T,\textbf{X}) + \mathcal{O}(Z^2) \right] \right).
\end{split}
\end{align}
We can now determine which of the fields corresponds to a source in the CFT and which describes a physical fluctuation of the CFT without sources
by requiring that the norm of the physical state is finite.

The norm of a spinor in curved space-time is given by
\begin{align}
	\braket{\Psi|\Psi} = \int_\Sigma d \Sigma \sqrt{g_\Sigma} ~ n_M  e_A^M  \bar{\Psi} \gamma^A \Psi,
\end{align}
where $n^M$ is a future-directed unit vector orthogonal to the spacelike Cauchy surface $\Sigma$,
$g_\Sigma$ is the determinant of the induced metric and $\bar{\Psi} = \Psi^\dagger \Gamma^0$.
Divergences of the norm may arise from contributions near the boundary.
Choosing a constant time slice, we obtain
\begin{align}
	\braket{\Psi|\Psi} \sim \int_{t=\text{const}} d Z ~ Z^{-d} \Psi^\dagger \Psi.
\end{align}
By plugging in the expansion \eqref{spinorasym} and requiring finiteness of the norm, we distinguish three cases:
\begin{itemize}
  \item If $m>\frac{1}{2}$, the Bessel functions of the first kind are normalizable and $\psi$ is a physical fluctuation on the boundary, while $\chi$ is a source.
  \item If $m<-\frac{1}{2}$, the Bessel functions of the second kind are normalizable and the role of $\psi$ and $\chi$ is interchanged.
\item In the window $-\frac{1}{2} < m < \frac{1}{2}$ both solutions are normalizable, and one must decide which field represents the source and which
represents a physical (source-free) fluctuation \cite{Klebanov:1999tb}.
\end{itemize}

A convenient way to characterize the physical field is in terms of the boundary condition it satisfies.  The possibilities are
\begin{align}
\label{AdSbcJapp}
& \hbox{\rm J-type or standard branch:} \quad\, \Gamma^Z \Psi \doteq - i \Psi \quad \text{as } Z \to 0 \\
\label{AdSbcYapp}
& \hbox{\rm Y-type or alternate branch:} \quad \Gamma^Z \Psi \doteq  i \Psi \quad \text{as } Z \to 0.
\end{align}
($\doteq$ means the coefficient of the leading $Z \rightarrow 0$ behavior of the two sides is the same.)
One can see this $Z \rightarrow 0$ behavior explicitly in (\ref{AdSoddextrapolate}), (\ref{AdSoddextrapolateAlternate}), (\ref{AdSevenextrapolate}), (\ref{AdSevenextrapolateAlternate}).
These boundary conditions have several desirable features.  They lead to a well-defined variational principle, after adding suitable surface terms to the Dirac action \cite{Henningson:1998cd,Henneaux:1998ch}.  Also, as discussed for example in \cite{Dias:2019fmz}, they imply a vanishing flux of momentum and fermion number through the boundary.

These bounds we have given are the equivalent of the Breitenlohner-Freedman bound for fermions.
We identify the scaling dimension of the fermion for the two possible branches as
\begin{align}
	\Delta = \pm m + \frac{d}{2},
\end{align}
where the plus sign is chosen for the J-type solutions and the minus sign for the Y-type solutions, respectively.
We note that the unitarity bound of a CFT with spin 1/2 fields \cite{Minwalla:1997ka}
\begin{align}
	\Delta \ge \frac{d-1}{2}
\end{align}
is always satisfied and is saturated if $m = \mp \frac{1}{2}$.

%%%%%%%%%%%%%%%%%%%%%%%%%
\section{Fermion representations in general dimensions\label{appendix:spinor}}
%%%%%%%%%%%%%%%%%%%%%%%%%
In this appendix we review some standard facts about fermion representations and mass terms in general dimensions.  To avoid repetition we adopt notation and
borrow results from Sohnius \cite{Sohnius:1985qm}, Section 14.1 and Appendix A.7.

First recall that reality (Majorana) and chirality (Weyl) conditions can be imposed in the following space-time dimensions.  Assuming one time dimension, and with $d$ denoting the total
number of space-time dimensions,\footnote{So in the body of this paper $d$ refers to the spacetime dimension of the CFT, while in this appendix it could refer to either bulk or boundary.
Also in this appendix $\psi$ could refer to either a bulk or boundary spinor.  Note that our Clifford algebra convention corresponds to $d_+ = 1$ in Sohnius.} we have (table 14.1 in Sohnius)
\begin{center}
\begin{tabular}{ll}
spinor & space-time dimension \\
\hline
Dirac & any $d$ \\
Majorana & $d = 0,1,2,3,4$ (mod 8) \\
chiral & $d$ even \\
Majorana--Weyl & $d = 2$ (mod 8)
\end{tabular}
\end{center}

Next, let's study how charge conjugation $\psi^C = X \psi^*$ acts on Dirac matrices.  The options for the matrix $X$ are
\begin{eqnarray*}
\begin{array}{ll}
\hbox{\rm AdS${}_{\rm even}$:} & \hbox{\rm  either $X = D$ or $X = \tilde{D}$, with} \\[2pt]
& \quad D^{-1} \Gamma_a D = - (\Gamma_a)^* \\
& \quad \tilde{D}^{-1} \Gamma_a \tilde{D} = + (\Gamma_a)^* \\[10pt]
\hbox{\rm AdS${}_{\rm odd}$:} & \hbox{\rm only $X = D$ is available, with} \\[2pt]
& \quad \hbox{\rm $D^{-1} \Gamma_a D = + (\Gamma_a)^*$ if $d = 1$ (mod 4)} \\
& \quad \hbox{\rm $D^{-1} \Gamma_a D = - (\Gamma_a)^*$ if $d = 3$ (mod 4)}
\end{array}
\end{eqnarray*}
(see Sohnius (A.60) and (A.65) in even dimensions and table A.1 in odd dimensions).  Then from table A.4, among the dimensions where a Majorana condition is possible, we have
\begin{eqnarray*}
&& X^{-1} (\Gamma_a)^* X = - \Gamma_a \qquad \hbox{\rm if $d = 2, 3, 4$ (mod 8)} \\
&& X^{-1} (\Gamma_a)^* X = + \Gamma_a \qquad \hbox{\rm if $d = 8, 9$ (mod 8)}
\end{eqnarray*}

Next, let's see if we can write a fermion mass term.  In any number of dimensions we can write a mass term for a Dirac spinor,
\begin{equation}
{\cal L}_{\rm Dirac} = m \bar{\psi} \psi,
\end{equation}
where $\bar{\psi} = \psi^\dagger A$.  A real spinor satisfies $\psi^C = \psi$, where the charge conjugate spinor $\psi^C = C A^T \psi^*$.  (See equations (14.8) and (14.9) in Sohnius; the possibility
of including $\Gamma_{d+1}$ in the definition of $\psi^C$ leads to the same conclusion.)  So for a Majorana spinor it would seem we can write a mass term
\begin{equation}
{\cal L}_{\rm Majorana} = m \bar{\psi} \psi = m \bar{\psi} \psi^C = m (\psi^*)^T A C A^T \psi^*.
\end{equation}
The matrix $A C A^T$ appearing in the mass term satisfies
\begin{equation}
(A C A^T)^T = A C^T A^T = \eta A C A^T,
\end{equation}
where $\eta = \pm 1$ is defined in Sohnius (A.61).  If $\eta = -1$ the matrix is antisymmetric and a Majorana mass term is possible, but if $\eta = + 1$ the matrix is symmetric and the mass term
vanishes by Fermi statistics.  Then from table A.3 in Sohnius we have, among the dimensions where a Majorana spinor is possible,
\begin{center}
\begin{tabular}{ll}
$d = 0,1$ (mod 8) & $\eta = +1$, no mass term allowed \\
$d = 2,3,4$ (mod 8) & $\eta = -1$, a Majorana mass is possible
\end{tabular}
\end{center}

Now let's specialize to even dimensions and include the possibility of chiral spinors.  The basic property, which we'll establish below, is that in $d = 4k$ dimensions charge conjugation flips the chirality of a spinor while in $d = 4k + 2$ dimensions
it leaves chirality unchanged.  This means that in $4k$ dimensions real spinors and chiral spinors are equivalent since we can define
\begin{equation}
\psi_{\rm Majorana} = \psi_{\rm chiral} + \psi_{\rm chiral}^C \,.
\end{equation}
From the discussion above a mass term for these spinors is possible if $d = 4$ (mod 8) but not if $d = 0$ (mod 8).  On the other hand if $d = 4k + 2$ then $\psi$ and $\psi^C$ have the same chirality and there is no way to write a
Lorentz-invariant mass term.  To summarize, the options for defining chiral or real (but not both!) spinors and writing mass terms are
\begin{center}
\begin{tabular}{cccccc}
dimension & chiral? & real? & equivalent? & chiral mass? & real mass? \\
\hline
$d = 0$ (mod 8) & yes & yes & yes & no & no \\
$d = 1$ (mod 8) & no & yes & -- & -- & no \\
$d = 2$ (mod 8) & yes & yes & no & no & yes \\
$d = 3$ (mod 8) & no & yes & -- & -- & yes \\
$d = 4$ (mod 8) & yes & yes & yes & yes & yes \\
$d = 5$ (mod 8) & no & no & -- & -- & -- \\
$d = 6$ (mod 8) & yes & no & -- & no & -- \\
$d = 7$ (mod 8) & no & no & -- & -- & --
\end{tabular}
\end{center}
Note that we're indicating whether it's possible to write a mass term for a single fermion.  In some cases when this is forbidden it's still possible to write a symplectic mass term for an even number of fermion
species \cite{figueroamajorana}.

The discussion can be summarized in the following table, where we've listed the possible spinors and underlined the dimensions in which a mass term is allowed.  For completeness we've included
Majorana-Weyl spinors, possible only if $d = 2$ (mod 8); a mass term is not allowed for these spinors since it's already forbidden for chiral spinors.
\begin{center}
\begin{tabular}{ll}
spinor & space-time dimension \\
\hline
Dirac & mass allowed in any $d$ \\
Majorana & $d = 0, 1, \underline{2}, \underline{3}, \underline{4}$ (mod 8) \\
chiral & $d = 0, 2, \underline{4}, 6$ (mod 8) \\
Majorana--Weyl & $d = 2$ (mod 8)
\end{tabular}
\end{center}

It only remains to establish the claim above, that charge conjugation flips the chirality of a spinor in $4k$ dimensions but leaves it invariant in $4k+2$.  Following Sohnius we denote
\begin{eqnarray}
&&\hbox{\rm charge-conjugate spinor} \quad \psi^C = X \psi^* \\
&&\hbox{\rm chirality operator} \quad \Gamma_{d+1} = \Gamma_1 \cdots \Gamma_d,
\end{eqnarray}
where either $X = D$ or $X = \tilde{D}$.
Consider a chiral spinor (an eigenstate of $\Gamma_{d+1}$) with $\Gamma_{d+1} \psi = \lambda \psi$ so that $\Gamma_{d+1}^* \psi^* = \lambda^* \psi^*$.  From Sohnius (A.60a) and
(A.65) note that
\begin{equation}
\Gamma_{d+1} X = X X^{-1} \Gamma_1 X X^{-1} \Gamma_2 \cdots \Gamma_d X = (\pm 1)^d X \Gamma_{d+1}^*
\end{equation}
($-$ for $D$, $+$ for $\tilde{D}$).  But since $d$ is even
\begin{equation}
\Gamma_{d+1} \psi^C = \Gamma_{d+1} X \psi^* = X \Gamma_{d+1}^* \psi^* = X \lambda^* \psi^* = \lambda^* \psi^C.
\end{equation}
So $\psi^C$ is also an eigenstate, but with eigenvalue $\lambda^*$.
From equation (A.61) and table A.2 in Sohnius we have $(\Gamma_{d+1})^2 = \beta \mathbb{1}$ where
\begin{equation}
\beta = \left\lbrace\begin{array}{ll}
-1 \quad & \hbox{\rm when $d = 0$ (mod 4)} \\
+1 \quad & \hbox{\rm when $d = 2$ (mod 4)}
\end{array}\right.
\end{equation}
So in $4k$ dimensions the possible eigenvalues of $\Gamma_{d+1}$ are $\pm i$.  Then $\lambda^* = - \lambda$ which means $\psi$ and $\psi^C$ have opposite chirality.
But in $4k+2$ dimensions the eigenvalues of $\Gamma_{d+1}$ are $\pm 1$.  Then $\lambda^* = \lambda$ so $\psi$ and $\psi^C$ have the same chirality.

%%%%%%%%%%%%%%%%%%%%%%%%%%
%\bibliographystyle{utphys}
%\bibliography{fermions}
\providecommand{\href}[2]{#2}\begingroup\raggedright\endgroup

\end{document}